\documentclass[preprint,12pt]{emulateapj}

\usepackage{apjfonts}
\usepackage{color}
\usepackage{natbib}
\usepackage{amssymb,amsmath}
\usepackage{graphicx}
\usepackage[colorlinks,urlcolor=red,citecolor=blue,linkcolor=cyan]{hyperref} 

\DeclareRobustCommand{\ion}[2]{%
\relax\ifmmode
\ifx\testbx\f@series
{\mathbf{#1\,\mathsc{#2}}}\else
{\mathrm{#1\,\mathsc{#2}}}\fi
\else\textup{#1\,{\mdseries\textsc{#2}}}%
\fi}

\shortauthors{Albrecht et al.} \shorttitle{CV\,Velorum}
\begin{document}

\title{The BANANA project.~V.~Misaligned and precessing stellar rotation axes
  in CV\,Velorum.$^{\star}$}

\author{
Simon Albrecht\altaffilmark{1}, 
Joshua N.\ Winn\altaffilmark{1},
Guillermo Torres\altaffilmark{2},
Daniel C.\ Fabrycky\altaffilmark{3},
Johny Setiawan\altaffilmark{4,5},\\
Micha\"{e}l Gillon\altaffilmark{6},
Emmanuel Jehin\altaffilmark{6}, 
Amaury Triaud\altaffilmark{1,7,8},
Didier Queloz\altaffilmark{7,9},
Ignas Snellen\altaffilmark{10},
Peter Eggleton\altaffilmark{11}
}

\altaffiltext{1}{Department of Physics, and Kavli Institute for
  Astrophysics and Space Research, Massachusetts Institute of
  Technology, Cambridge, MA 02139, USA}

\altaffiltext{2}{Harvard-Smithsonian Center for Astrophysics,
  Cambridge, MA 02138, USA}

\altaffiltext{3}{Department of Astronomy and Astrophysics, University
  of California, Santa Cruz, Santa Cruz, CA 95064, USA}

\altaffiltext{4}{Max-Planck-Institut f\"{u}r Astronomie,
  K\"{o}nigstuhl 17, 69117 Heidelberg, Germany}

\altaffiltext{5}{Embassy of the Republic of Indonesia, Lehrter Str.\
  16-17, 10557 Berlin, Germany}

\altaffiltext{6}{Institut d'Astrophysique et de G\'{e}ophysique,
  Universit\'{e} de Li\`{e}ge, Allée du 6 Ao\^{u}t, 17, Bat.\ B5C,
  Li\`{e}ge 1, Belgium}

\altaffiltext{7}{Observatoire Astronomique de l\'{U}niversité de
  Gen\`{e}ve, Chemin des Maillettes, 51, Sauverny, CH-1290,
  Switzerland}

\altaffiltext{8}{Fellow of the Swiss national science foundation}

\altaffiltext{9}{Astrophysics Group, Cavendish Laboratory, JJ Thomson
  Avenue, Cambridge CB3 0HE, United Kingdom}

\altaffiltext{10}{Leiden Observatory, Leiden University, Niels Bohrweg
  2, 2333 CA Leiden, the Netherlands}

\altaffiltext{11}{Lawrence Livermore National Laboratory, 7000 East
  Ave, Livermore, CA 94551, USA}

\altaffiltext{$\star$}{Based on observations made with ESOs $2.2$\,m
  Telescopes at the La~Silla~Paranal~Observatory under programme ID
  084.C-1008 and under MPIA guaranteed time. }

\begin{abstract}

  As part of the BANANA project (Binaries Are Not Always Neatly
  Aligned), we have found that the eclipsing binary CV\,Velorum has
  misaligned rotation axes. Based on our analysis of the
  Rossiter-McLaughlin effect, we find sky-projected spin-orbit angles
  of $\beta_{\rm p} = -52\pm6^{\circ}$ and $\beta_{\rm s}=
  3\pm7^{\circ}$ for the primary and secondary stars (B2.5V~$+$~B2.5V,
  $P=6.9$~d).  We combine this information with several measurements
  of changing projected stellar rotation speeds ($v \sin i_{\star}$)
  over the last $30$ years, leading to a model in which the primary
  star's obliquity is $\approx65^{\circ}$, and its spin axis
  precesses around the total angular momentum vector with a period of
  about $140$ years. The geometry of the secondary star is less clear,
  although a significant obliquity is also implicated by the observed
  time variations in the $v \sin i_{\star}$. By integrating the
  secular tidal evolution equations backward in time, we find that the
  system could have evolved from a state of even stronger misalignment
  similar to DI\, Herculis, a younger but otherwise comparable binary.

\end{abstract}


\section{Introduction}
\label{sec:intro}

\setcounter{footnote}{0} 

Stellar obliquities (spin-orbit angles) have been measured in only a
handful of binary stars. \citep[see][for a
compilation]{albrecht2011}. The case of DI\,Herculis, in which both
stars have large obliquities \citep{albrecht2009}, has taught us that
it is risky to assume that the orbital and spin axes are aligned.  For
one thing, misalignment can influence the observed stellar parameters;
for example, the rotation axes may precess, producing time variations
in the sky-projected rotation speeds
\citep{albrecht2009,reisenberger1989}. Precession would also produce
small changes in the orbital inclination, and therefore changes in any
eclipse signals. Misalignment also influences the rate of apsidal
precession, as predicted by \cite{shakura1985} for the case of DI\,Her
and confirmed by \cite{albrecht2009}, and as is suspected to be the
case for AS\,Cam \citep{pavlovski2011}.

Furthermore, measurements of obliquities should be helpful in
constraining the formation and evolution of binary systems.  Larger
obliquities might indicate that a third star on a wide, inclined orbit
gave rise to Kozai cycles in the close pair during which the close
pair's orbital eccentricity and inclination oscillate
\citep{mazeh1979,eggleton2001,fabrycky2007,naoz2013}.  A third body on
an inclined orbit might also cause orbital precession of the inner
orbit around the total angular momentum, again creating large opening
angles between the stellar rotation and the orbit of the close pair
\citep{eggleton2001}. There are also other mechanisms which can lead
to apparent misalignments. For example \cite{rogers2012,rogers2013}
suggest that for stars with an outer radiative layer, large angles
between the convective core and the outer layer may be created by
internal gravity waves.

The aim of the BANANA project (Binaries Are Not Always Neatly Aligned)
is to measure obliquities in close binaries and thereby constrain
theories of binary formation and evolution. We refer the reader to
\cite{albrecht2011} for a listing of different techniques to measure
or constrain obliquities. \cite{triaud2013}, \cite{lehman2013},
\cite{philippov2013}, and \cite{zhou2013} have also presented new
obliquity measurements in some binary star systems. This paper is
about the CV\,Vel system, the fifth BANANA system. Previous papers
have examined the V1143\,Cyg, DI\,Her, NY\,Cep, and EP\,Cru systems
\citep[][Papers
I-IV]{albrecht2007,albrecht2009,albrecht2011,albrecht2013}.

\paragraph{CV\,Velorum} 
CV\,Vel was first described by \cite{houten1950}.  A spectroscopic
orbit and a light curve were obtained by \cite{feast1954} and
\cite{gaposchkin1955}.  Two decades later, \cite{andersen1975}
analyzed the system in detail. Together with four Str\"{o}mgen light
curves obtained and analyzed by \cite{clausen1977}, this has allowed
the absolute dimensions of the system to be known with an accuracy of
about 1\%.  More recently \cite{yakut2007} conducted another study of
the system, finding that the stars in the system belong to the class
of slowly pulsating B stars \citep{waelkens1991,decat2002}.
Table~\ref{tab:cvvel} summarizes the basic data.

\begin{table}[t]
  \caption{General data on CV\,Velorum}
  \label{tab:cvvel}
  \begin{center}
    \smallskip
    \begin{tabular}{l l l}
      \hline
      \hline
      \noalign{\smallskip}
      HD number & 77464  &  \\ 
      HIP number & 44245 &  \\ 
      \noalign{\smallskip}
      \hline 
      \noalign{\smallskip}
      R.A.$_{\rm J2000}$ & $09^{\rm h}00^{\rm m}38^{\rm s}$&\tablenotemark{$^{\rm a}$} \\
      Dec.$_{\rm J2000}$ & $-51^{\circ}33^{\prime}20^{\prime\prime}$&\tablenotemark{$^{\rm a}$}  \\
      Distance  & $439_{60} ^{82}$ pc &\tablenotemark{$^{\rm a}$}\\
      V$_{\mbox{max}}$  & $6.69$\,mag& \tablenotemark{$^{\rm b}$} \\
      Sp.\ Type  & B2.5V+ B2.5V &\tablenotemark{$^{\rm c}$} \\
      Orbital period  & $6\fd889$&\tablenotemark{$^{\rm b}$}\\
      Eccentricity  & $0$&\tablenotemark{$^{\rm b}$} \\
      Primary mass, $M_{\rm p}$ & $6.086(44)~M_{\odot}$&\tablenotemark{$^{\rm d}$} \\
      Secondary mass, $M_{\rm s}$ & $5.982(35)~M_{\odot}$&\tablenotemark{$^{\rm d}$} \\
      Primary radius, $R_{\rm p}$ & $4.089(36)~R_{\odot}$&\tablenotemark{$^{\rm d}$} \\
      Secondary radius, $R_{\rm s}$ & $3.950(36)~R_{\odot}$&\tablenotemark{$^{\rm d}$} \\
      Primary effective temperature ,$T_{\rm eff\,p}$ & $18\,100(500)$~K&\tablenotemark{$^{\rm d}$} \\
      Secondary effective temperature, $T_{\rm eff\,s}$ & $17\,900(500)$~K&\tablenotemark{$^{\rm d}$}  \\
      Age & $40\cdot10^{6}$\,yr&\tablenotemark{$^{\rm e}$} \\
      \noalign{\smallskip}
      \noalign{\smallskip}
      \hline
      \noalign{\smallskip}
      \noalign{\smallskip}
      $^{\rm a}$ \cite{vanleeuwen2007}\\
      $^{\rm b}$ \cite{clausen1977}\\
      $^{\rm c}$ \cite{andersen1975}\\
      $^{\rm d}$ \cite{torres2010}\\
      $^{\rm e}$ \cite{yakut2007}\\    
    \end{tabular}
  \end{center} 
\end{table}

One reason why this binary was selected for BANANA was the
disagreement in the measured projected stellar rotation speeds
measured by \cite{andersen1975} and \cite{yakut2007}.
\cite{andersen1975} found $v \sin i_{\star}=28\pm3$~km\,s$^{-1}$ for
both stars. Employing data obtained nearly $30$ years later,
\cite{yakut2007} found $v \sin i_{\rm p} = 19\pm1$~km\,s$^{-1}$ and $v
\sin i_{\rm s} = 31\pm2$~km\,s$^{-1}$, indicating a significantly
lower $v \sin i_{\star}$ for the primary star.  This could indicate
that the stellar rotation axes are misaligned and precessing around
the total angular momentum vector, as has been observed for
DI\,Herculis \citep{albrecht2009}.

This paper is structured as follows. In the next section we describe
the observations. In Section~\ref{sec:analysis} we describe our
analysis method and results. Then, in Section~\ref{sec:cvvel_rot}, we
discuss our findings in the framework of tidal evolution. We end in
Section~\ref{sec:summary} with a summary of our conclusions.

\section{Spectroscopic and Photometric Data}
\label{sec:data}

\paragraph{Spectroscopic observations} 

We observed CV\,Vel with the FEROS spectrograph \citep{kaufer1999} on
the $2.2$~m telescope at ESO's La\,Silla observatory. We obtained
$100$ observations on multiple nights between December 2009 and March
2011 with a typical integration time of 5-10\,min.  We obtained $26$
spectra during $3$ primary eclipses, $30$ spectra during $4$ secondary
eclipses, and another $44$ spectra outside of eclipses. In addition we
observed the system with the {\it CORALIE} spectrograph at the Euler
Telescope at La\,Silla. Two spectra were obtained in 2010, and another
4 spectra were obtained in spring 2013. The {\it CORALIE} observations
were made near quadrature.

In all cases, we used the software installed on the observatory
computers to reduce the raw 2-d CCD images and to obtain stellar flux
density as a function of wavelength. The uncertainty in the wavelength
solution leads to a velocity uncertainty of a few m\,s$^{-1}$, which
is negligible for our purposes. The resulting spectra have a
resolution of $\approx$$50\,000$ around $4500$~\AA{}, the wavelength
region relevant to our analysis. We corrected for the radial velocity
(RV) of the observatory, performed initial flat fielding with the
nightly blaze function, and flagged and omitted bad pixels.

While this work is mainly based on our new spectra, some parts of our
analysis also make use of the spectra obtained by
\cite{yakut2007}. Those earlier spectra help to establish the time
evolution of the stellar rotation axes over the last few years.

\begin{figure}
  \begin{center}
    \includegraphics[width=8.cm]{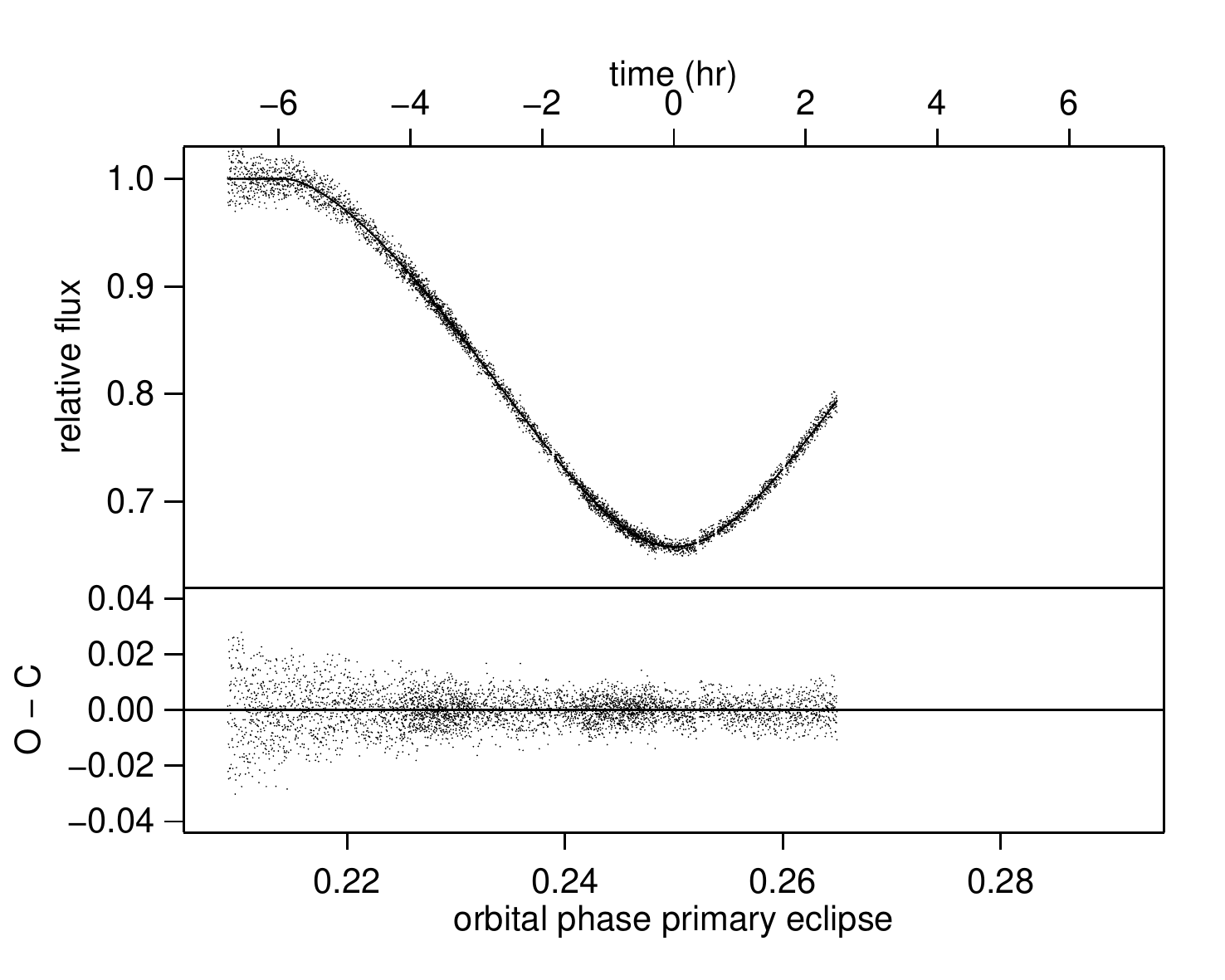}
    \includegraphics[width=8.cm]{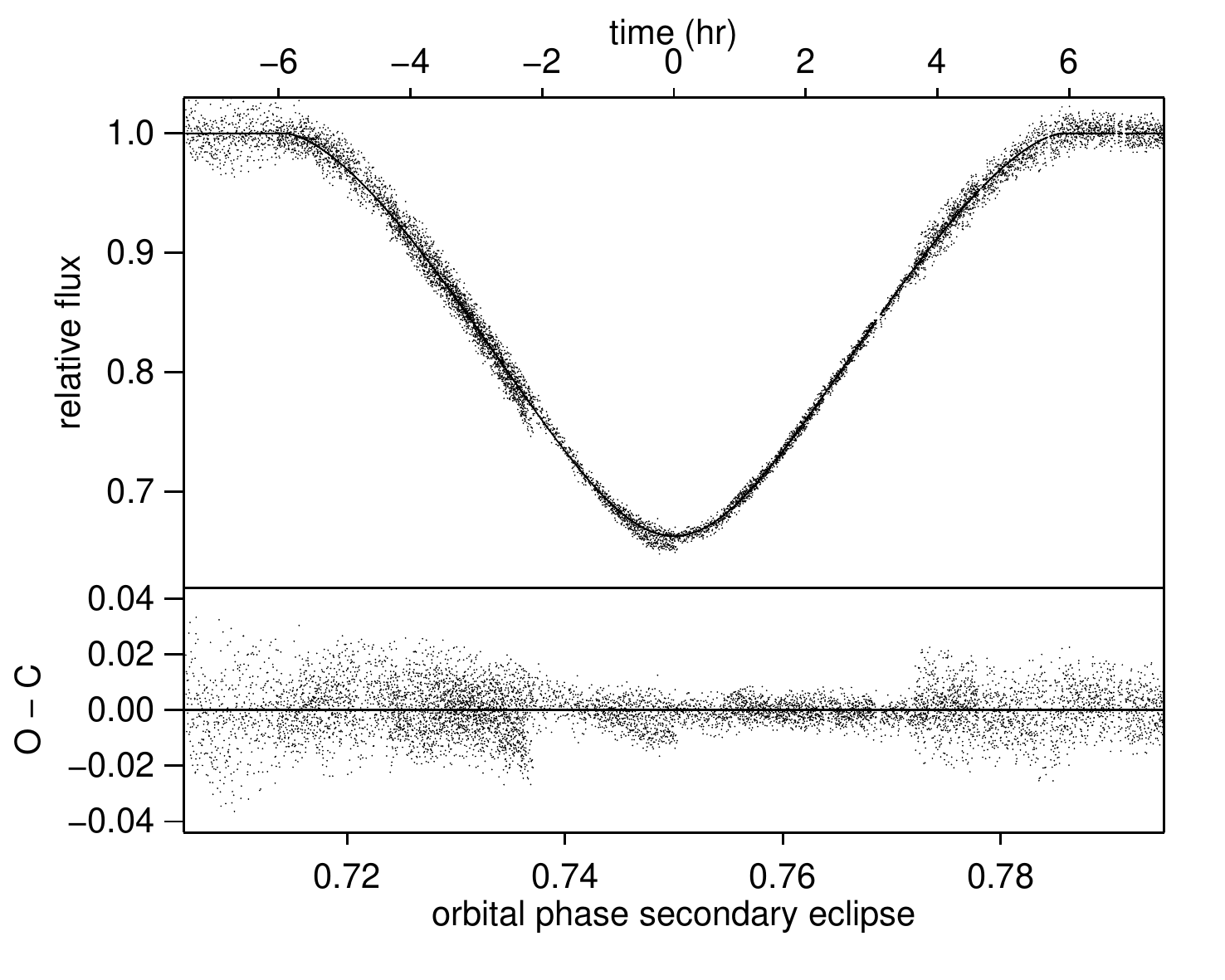}
    \caption {\label{fig:cvvel_photometry} {\bf Photometry of
        CV\,Vel.} {\it Top.}---Loss of light during primary eclipses,
      as observed with TRAPPIST.  The ``O-C''
      (observed~$-$~calculated) subpanel shows the residuals between
      the data and the best-fitting model. The model fits were also
      based on data from Fig.~1 of \cite{clausen1977}, which is not
      shown here.  {\it Bottom.}---Same, for secondary
      eclipses. In these plots phase 0 is offset by $0.25$ from the time of
      primary mid eclipse. }
  \end{center}
\end{figure}

\paragraph{Photometric  observations} 

To establish a modern eclipse ephemeris we obtained new photometric
data.  CV\,Vel was observed with the 0.6m TRAPPIST telescope in the I
and z bandpasses \citep{gillion2011} in La\,Silla\footnote{\tt
  http://www.astro.ulg.ac.be/Sci/Trappist}.  We observed the system
during several eclipses from November 2010 to January 2011. Since the
eclipses last nearly $12$~hr, only a portion of an eclipse can be
observed in a single night. We observed primary eclipses on three
different nights, and secondary eclipses on six different nights.
This gave full coverage of all phases of the secondary eclipse, and
coverage of about three-quarters of the primary eclipse (see
Figure~\ref{fig:cvvel_photometry}). We also include the Str\"{o}mgen
photometry obtained by \cite{clausen1977} in our study. 

\begin{figure}
  \begin{center} \includegraphics[width=9.cm]{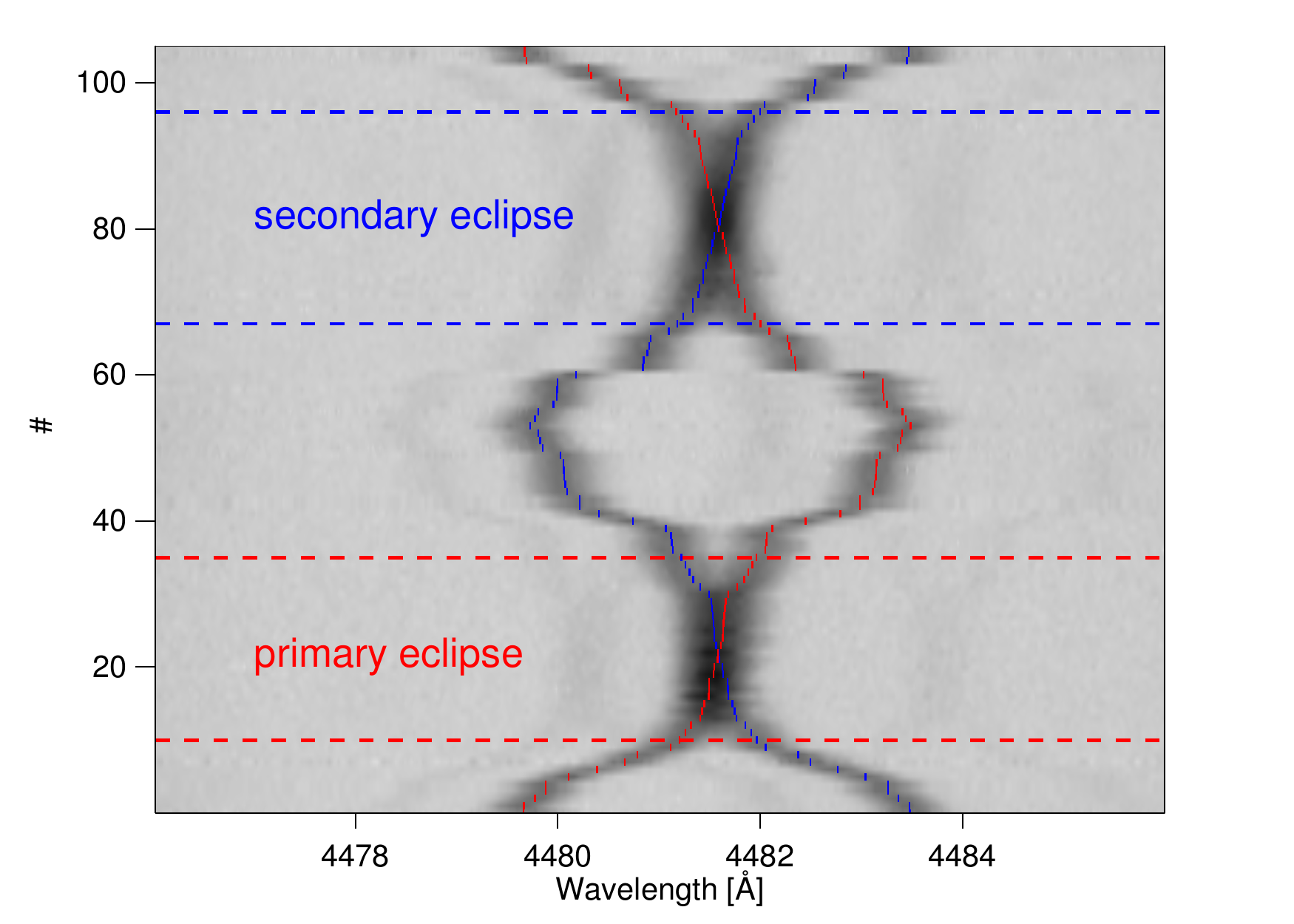}
    \caption {\label{fig:cvvel_grayscale} {\bf Spectroscopic
        observations of CV\,Vel.} Grayscale depiction of the spectra
      obtained outside eclipses and during both eclipses. The
      observations are sorted by orbital phase.  The horizontal dashed
      lines indicate the begin and end of the primary (red) and
      secondary (blue) eclipses. The dark bands represent absorption
      lines; the darkest band is the \ion{Mg}{ii} line. One can also
      see the weaker \ion{AI}{iii} and \ion{S}{ii} lines, at shorter
      and longer wavelengths, respectively. The small vertical blue
      and red lines indicate the calculated wavelength position of the
      central \ion{Mg}{ii} line, as expected from orbital motion. The
      lines overlap at times of eclipses. The discontinuities arise
      because of uneven coverage in orbital phase.}
  \end{center}
\end{figure} 

\section{Analysis}
\label{sec:analysis}

To measure the sky-projected obliquities, we take advantage of the
Rossiter-McLaughlin (RM) effect, which occurs during eclipses.  Below
we describe our general approach to analyzing the RM effect.
Section~\ref{sec:cvvel} describes some factors specific to the case of
CV\,Vel.  Section~\ref{sec:cvvel_results} presents the results.

\paragraph{Model} Our approach is similar to that described in
Papers~I--IV, where it is described in more detail. The projected
obliquity of stellar rotation axes can be derived from the
deformations of stellar absorption lines during eclipses, when parts
of the rotating photospheres are blocked from view, as the exact shape
of the deformations depend on the geometry of the eclipse.

We simulate spectra containing light from two stars. The simulated
spectra are then compared to the observed spectra, and the model
parameters are adjusted to provide the best fit. Our model includes
the orbital motion of both stars, and the broadening of the absorption
lines due to rotation, turbulent velocities, and the point-spread
function of the spectrograph (PSF). For observations made during
eclipses, the code only integrates the light from the exposed portions
of the stellar disks. The resulting master absorption line (which we
will call the ``kernel'') is then convolved with a line list which we
obtain from the Vienna Atomic Line Database
\citep[VALD;][]{kupka1999}. The lines are shifted in wavelength space
according to their orbital radial velocity, and weighed by the
relative light contribution from the respective stars. The model is
specified by a number of parameters.

\begin{figure}
  \begin{center}
    \includegraphics[width=9.cm]{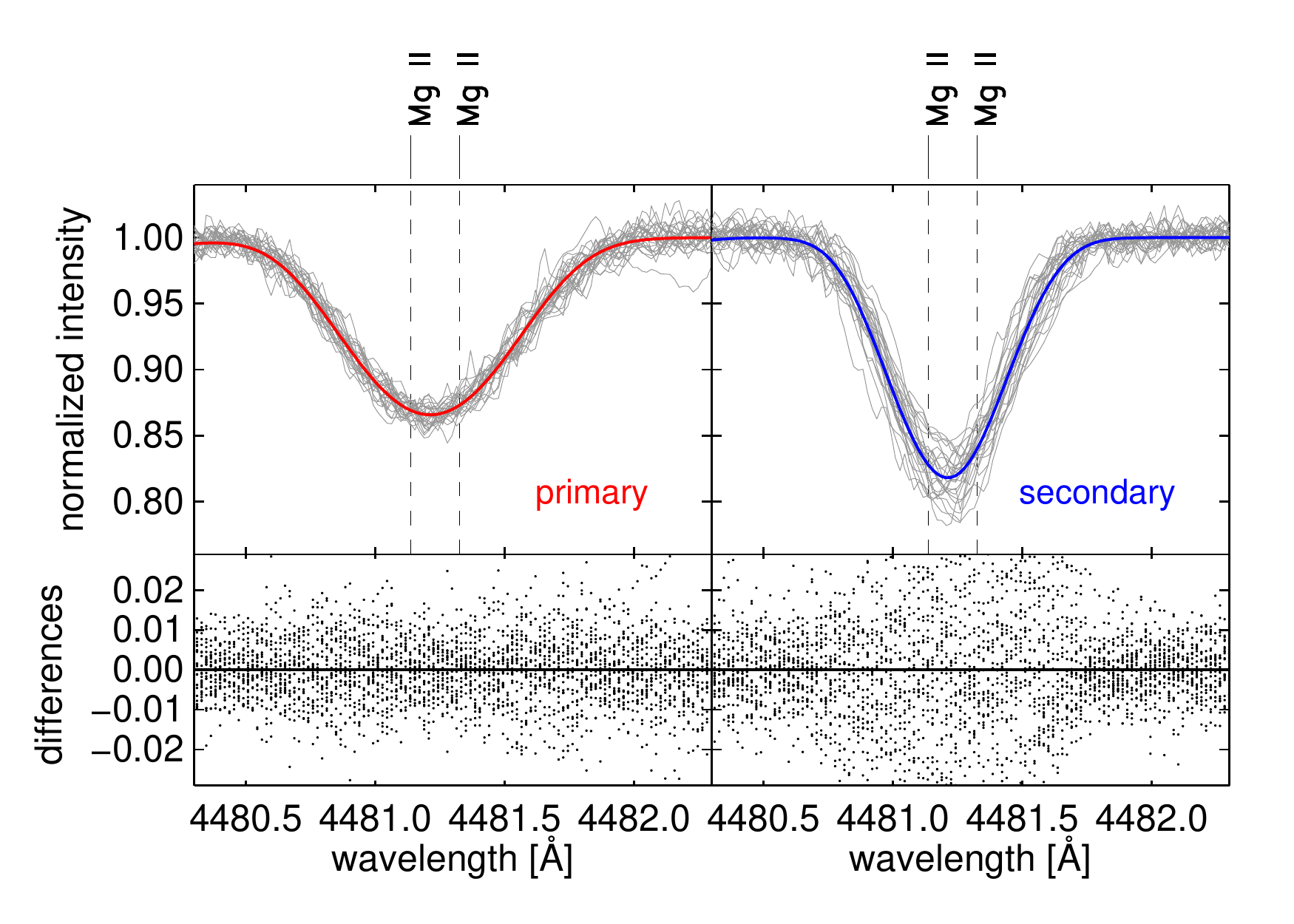}
    \caption {\label{fig:cvvel_lines_yakut} {\bf Absorption lines of
        CV\,Vel in 2001/2002.} {\it Left.}---The lines of the primary,
      in the region spanning the \ion{Mg}{ii} line. The best-fitting
      model for the secondary lines has been subtracted. Thin gray
      lines show all the out-of-eclipse observations. The red line
      shows our model for the primary lines. The lower panel shows the
      difference between individual observations and the mean
      line. {\it Right.}---The lines of the secondary, after
      subtracting the model of the primary lines. The blue line shows
      our model for the secondary. The pulsations of the secondary
      cause a larger scatter in the residuals. These spectra were
      obtained in December 2001 and January 2002 with the {\it
        CORALIE} spectrograph by \cite{yakut2007}. Note that
      \citep{yakut2007} mislabeled the primary as the secondary, and
      vice versa; here we have labeled the spectra correctly. 
        See also \citep{yakut2014}.}
  \end{center}
\end{figure}

\begin{figure}
  \begin{center}
    \includegraphics[width=9.cm]{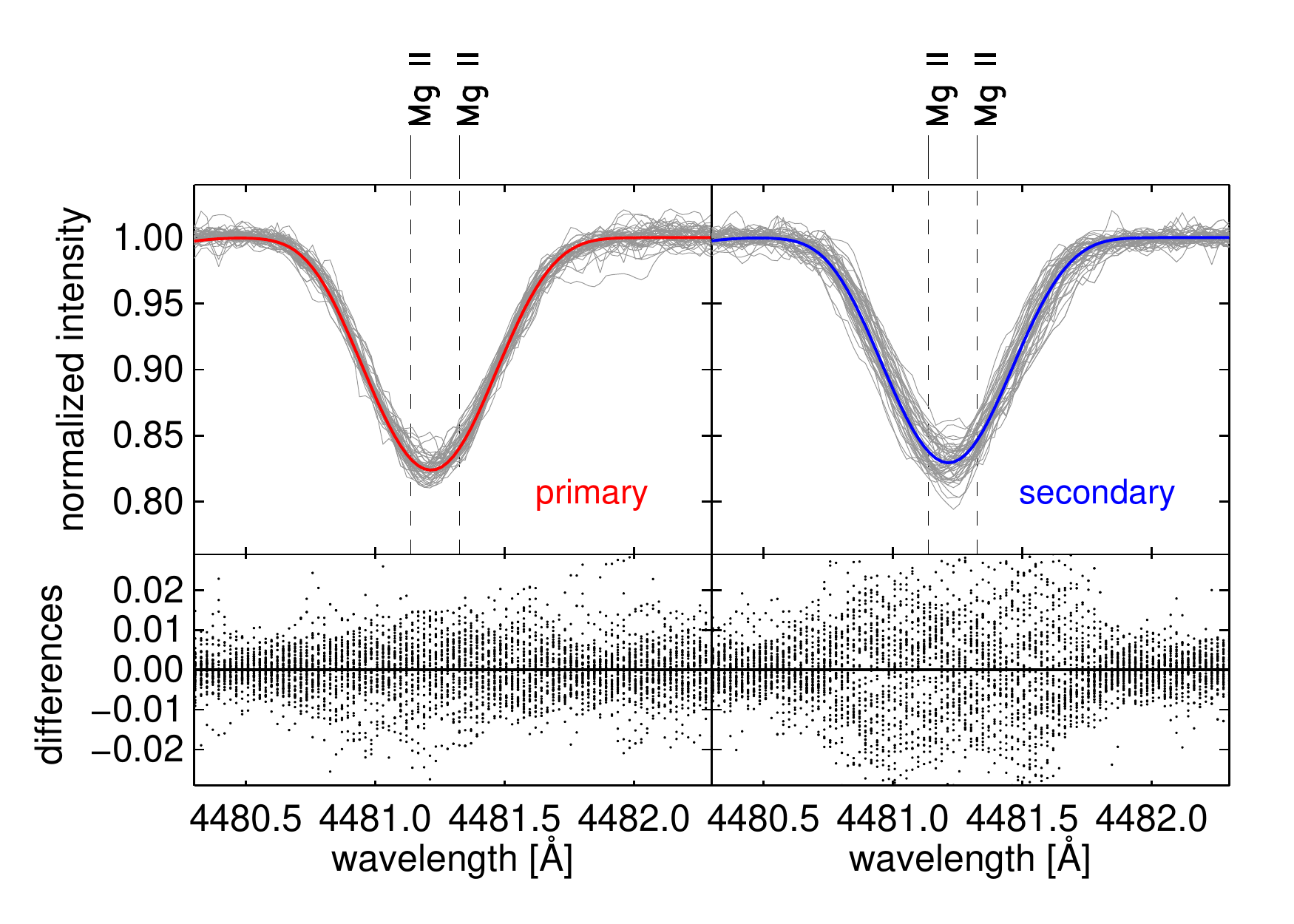}
    \caption {\label{fig:cvvel_lines} {\bf Absorption lines of CV\,Vel
        in 2009/2010.} Same as Fig.~\ref{fig:cvvel_lines_yakut} but
      for our FEROS and {\it CORALIE} spectra obtained outside of
      eclipses. Comparison to Fig.~\ref{fig:cvvel_lines_yakut} reveals
      that the primary absorption lines became significantly narrower
      between 2001/2002 and 2009/2010.}
  \end{center}
\end{figure}

\paragraph{Model parameters} The orbit is specified by the
eccentricity ($e$), argument of periastron ($\omega$), inclination
($i_o$), period ($P$), and RV semi-amplitudes of the primary and
secondary stars ($K_{\rm p}$ and $K_{\rm s}$). The position of the
stars on their orbits, and therefore the times of eclipses, are
defined by a particular epoch of primary mid-eclipse ($T_{\rm min,
  I}$). In addition, additive velocity offsets ($\gamma_{\rm i}$) are
needed.\footnote{We use one velocity offset for each star. Due to
  subtle factors specific to each star the $\gamma_{\rm i}$ can differ
  from the barycentric velocity of the system, and they can also
  differ between the stars.  Differences in gravitational redshift,
  line blending, and stellar surface flows could cause such shifts.}

To calculate the duration of eclipses and the loss of light, we need
to specify the fractional radius ($r \equiv R/a$, where $a$ is the
orbital semimajor axis) and quadratic limb darkening parameters ($u_1$
and $u_2$) for each star, as well as the light ratio between the two
stars ($L_{\rm s}/L_{\rm p}$) at the wavelength of interest.

\begin{figure}
  \begin{center}
    \includegraphics[width=8.9cm]{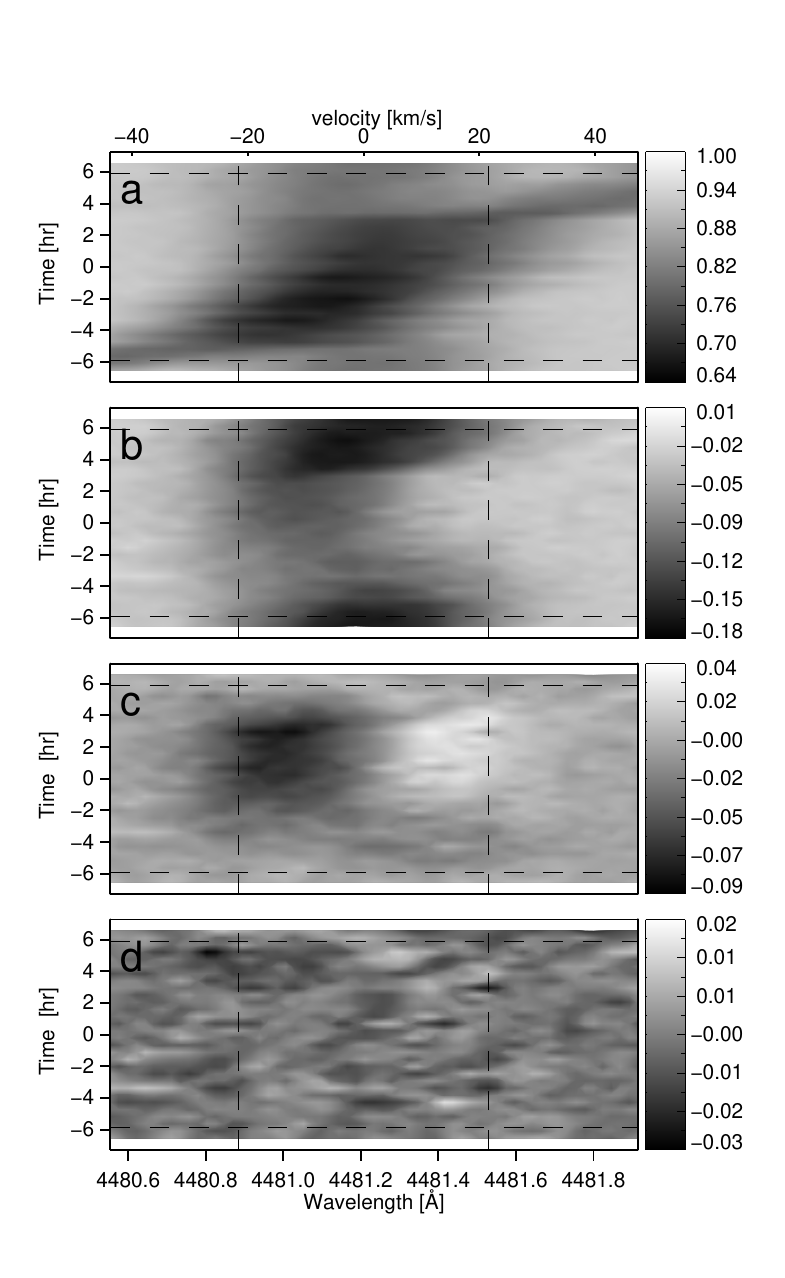}
    \caption {\label{fig:cvvel_primary_eclipse_grayscale} {\bf
        Observations of CV\,Vel during primary eclipse.} (a) Grayscale
      depiction of the time dependence of the \ion{Mg}{ii} lines of
      both stars, obtained throughout primary eclipses. The spectra
      have been shifted into the rest frame of the primary.
      Horizontal dashed lines mark the approximate boundaries of the
      primary lines, and vertical dashed lines mark the start and end
      of the eclipse.  (b) After subtracting the secondary lines,
      based on the best-fitting model (See also
      Figure~\ref{fig:cvvel_primary_eclipse_lines}).  (c) After
      further subtracting a model of the primary lines which does not
      account for the RM effect, but only the light loss during
      eclipse. This exposes the deformations due to the RM effect.
      Darkness indicates a deeper absorption line, lightness indicates
      a shallower depth than expected in the zero-RM transit
      model. Throughout most of the primary eclipse, the blueshifted
      side of the absorption line is deeper, indicating that the
      companion is almost exclusively eclipsing the receding half of
      the primary star. From this we can conclude that the primary
      rotation axis and the orbital axis are misaligned. (d) After
      subtracting a model including the RM effect.}
  \end{center}
\end{figure}

\begin{figure}
  \begin{center} \includegraphics[width=8.9cm]{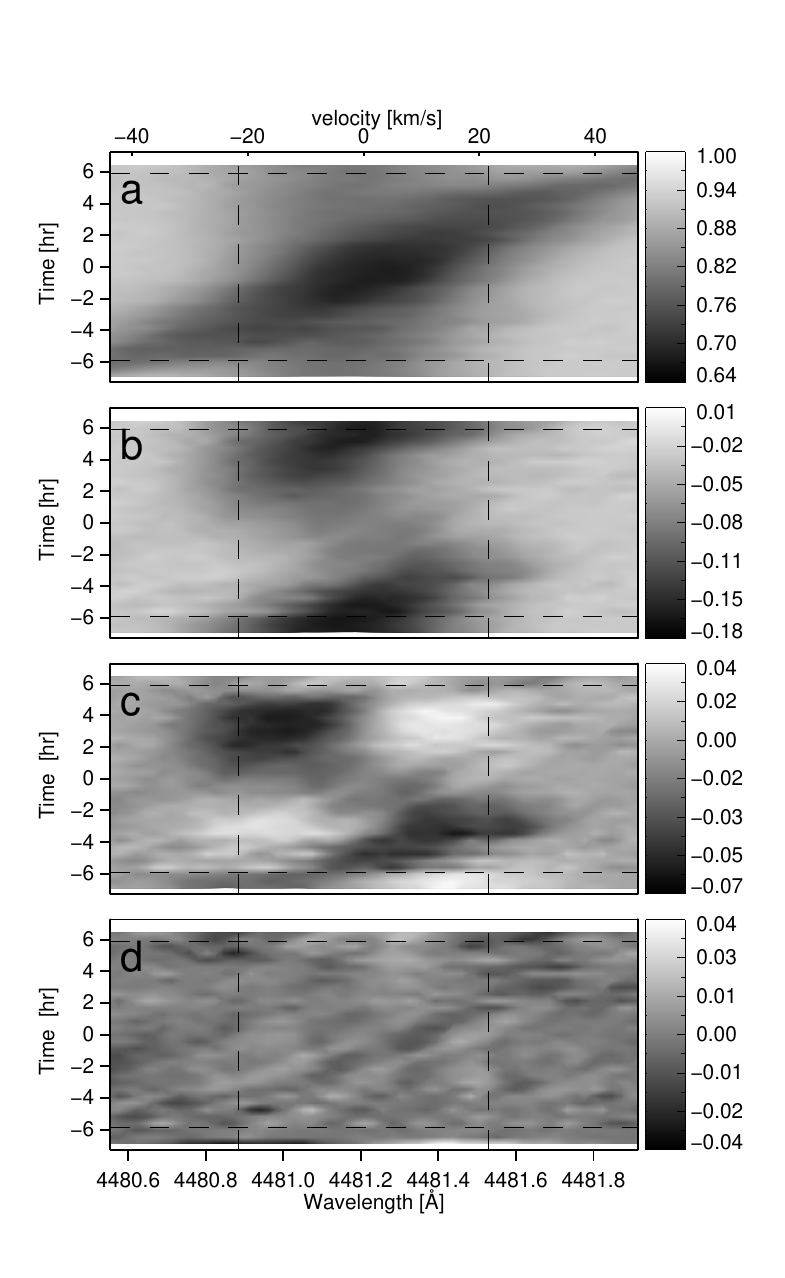}
    \caption {\label{fig:cvvel_secondary_eclipse_grayscale} {\bf
        Observations of CV\,Vel during secondary eclipse.} Similar to
      Figure~\ref{fig:cvvel_primary_eclipse_grayscale}, but for
      spectra obtained during secondary eclipses. In contrast to the
      deformations observed during primary eclipse, here the RM effect
      is antisymmetric in time and covers the full $v \sin i_{\star}$
      range, indicating alignment between the rotational and orbital
      axes on the sky.}
  \end{center}
\end{figure}

The kernel depends on various broadening mechanisms.  Assuming uniform
rotation, the rotational broadening is specified by $v \sin
i_{\star}$. Turbulent velocities of the stellar surfaces are described
with the micro-macro turbulence model of \cite{gray2005}. For this
model two parameters are required: the Gaussian width of the
macroturbulence ($\zeta_{\rm i}$); and the microturbulence parameter,
which is degenerate with the width of the spectrograph PSF.  We
specify the sky-projected spin-orbit angles ($\beta_{\rm p}$ and
$\beta_{\rm s}$) using the coordinate system (and sign convention) of
\citet{hosokawa1953}.

\paragraph{Normalization} To take into account the uncertainties due
to imperfect continuum normalization, we add 2 free parameters for
each spectrum, to model any residual slope of the continuum as a
linear function of wavelength. The parameters of the linear function
are optimized (in a separate minimization) every time a set of global
parameters are evaluated. This process is similar to the ``Hyperplane
Least Squares'' method that was described by \cite{bakos2010} and used
in the context of eclipses in double star systems by
\cite{albrecht2013}.

\begin{figure*}
  \begin{center}
    \includegraphics[width=15.5cm]{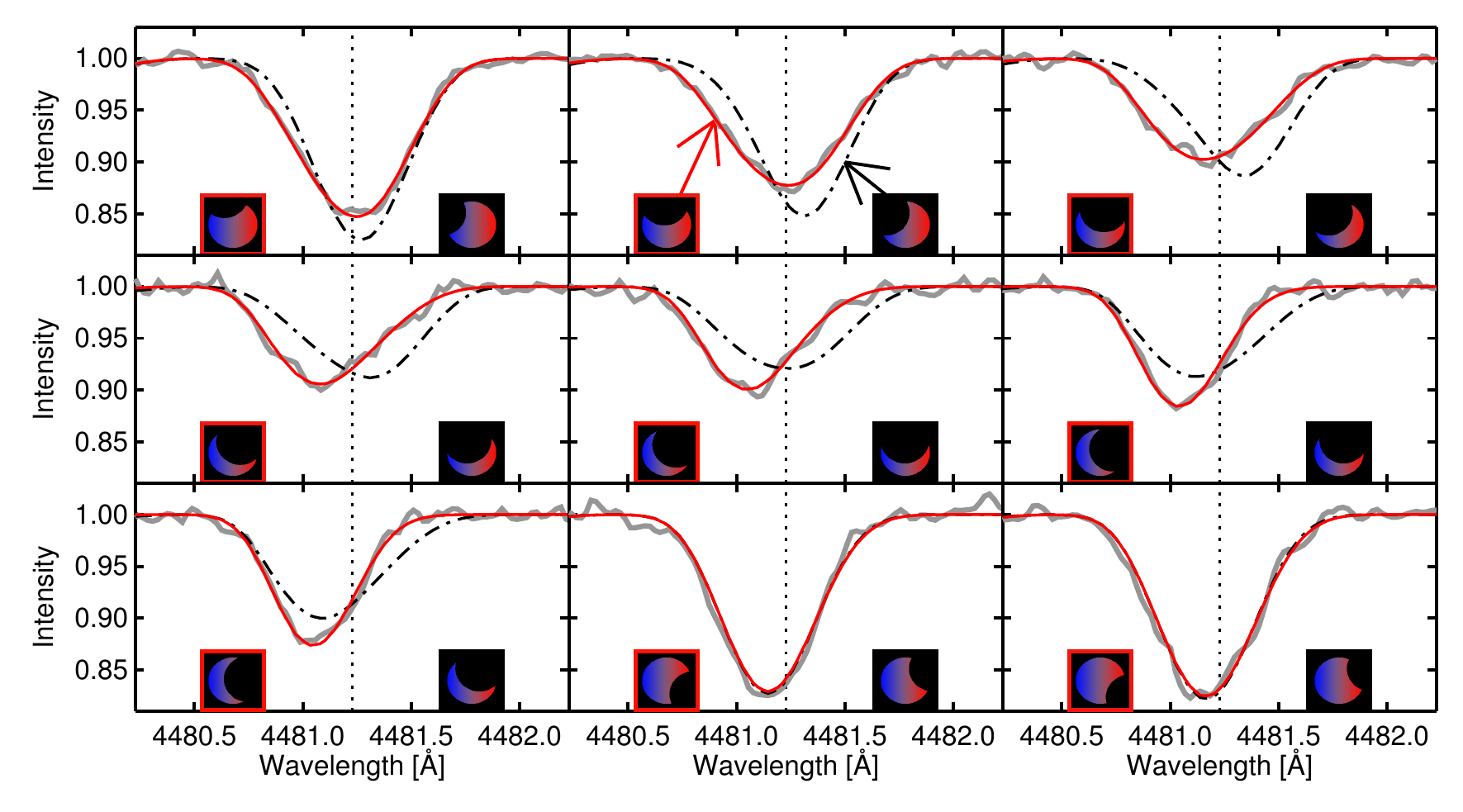}
    \caption {\label{fig:cvvel_primary_eclipse_lines} {\bf Absorption
        lines of CV\,Vel during primary eclipse.} Each panel shows the
      \ion{Mg}{ii} line of the primary star for a particular eclipse
      phase. The solid gray lines indicate the obtained spectra after
      subtraction of the best fitting model for the secondary and
      shifting the system in the restframe of the primary. The central
      wavelength of the \ion{Mg}{ii} line is indicated by the dashed
      line at $4481.228$~\AA{}. These nine panels show a subset of the
      observation presented in
      Figure~\ref{fig:cvvel_primary_eclipse_grayscale} and are
      presented in a way equivalent to panel (b). We also show our best
      fitting primary model as red solid line and as black dash-dotted
      line a model assuming co-aligned stellar and orbital axes,
      representing the data poorly. Each panel has two insets
      depicting the projected rotational velocity of the uncovered
      surface of the primary star, with blue and red indicating
      approaching and receding velocities, respectively. The left
      insets show the model with the misaligned stellar axis, and the
      right insets show the same for the co-aligned model.}
  \end{center}
\end{figure*}

\paragraph{Parameter estimation} To obtain parameter uncertainties we
used a Markov Chain Monte Carlo (MCMC) code. Our stepping parameters
were as listed above, except that instead of $i_o$ we stepped in $\cos
i_o$, and for the eccentricity parameters we stepped in $\sqrt{e} \cos
\omega$ and $\sqrt{e} \sin \omega$. The chains consisted of $0.5$
million calculations of $\chi^2$. The results reported below are the
median values of the posterior distribution, and the uncertainty
intervals are the values which exclude $15.85$~\% of the values at
each extreme of the posterior and encompass $68.3$~\% of the posterior
probability.

\subsection{Implementation of the model for CV\,Vel}
\label{sec:cvvel}

For CV\,Vel we focused on the \ion{Mg}{ii} line at $4481$~\AA{}, as
this line is relatively deep and broadened mainly by stellar
rotation. The projected stellar rotation speeds of both stars are
small (Table~\ref{tab:cvvel_results}). This means that the
\ion{Mg}{ii} line is well separated from the pressure-broadened
\ion{He}{i} line at $4471$~\AA{}, simplifying our analysis compared to
Papers~II--IV. In addition to the \ion{Mg}{ii} line, an \ion{Al}{iii}
doublet and a \ion{S}{ii} line are present in our spectral window from
$4476$~\AA{} to $4486$~\AA{}. Thus we include these lines in our
model.  Figure~\ref{fig:cvvel_grayscale} shows a grayscale
representation of all our observations in this wavelength range.

\cite{yakut2007} reported that the two members of the CV\,Vel system
belong to the class of slowly pulsating B stars \citep{waelkens1991}.
Using {\it CORALIE} spectra from December 2001 and January 2002, they
observed pulsations in both stars, with the pulsation amplitude for
the primary being larger than for the secondary (see their Figure
5c). We reanalyzed their spectra and found that they had mislabeled
the primary as the secondary, and vice versa. It is the secondary star
which showed the larger pulsations in their {\it CORALIE} spectra. In
addition, the values of $v \sin i_\star$ quoted by \citet{yakut2007}
were assigned to the wrong stars \citep{yakut2014}. In fact,
their measurement of $v \sin i_\star = 31\pm2$~km\,s$^{-1}$ belongs to
the primary, and their measurement of $v\sin i_\star=
19\pm1$~km\,s$^{-1}$ belongs to the secondary. See
Figure~\ref{fig:cvvel_lines_yakut}. Our observations took place about
a decade later. We also observed large pulsations in the spectra of
the secondary star (Figure~\ref{fig:cvvel_lines}). Here we describe
how we dealt with the pulsations while determining the projected
obliquIties.

The pulsation period is a few days. The out-of-eclipse observations
spanned many months, averaging over many pulsation periods. Thus the
pulsations likely introduce additional scatter into the derived
orbital parameters, but probably do not introduce large systematic
biases in the results. The situation is different for observations
taken during eclipses. Over the relatively short timespan of an
eclipse, the spectral-line deformation due to pulsation is nearly
static or changes coherently, and can introduce biases in the
parameters which are extracted from eclipse data. This is true not
only for the parameters of the pulsating star, but also for the
parameters of the companion star, since the light from both stars is
modeled simultaneously. Given the S/N of our spectra, the pulsations
of the primary star are too small to be a concern, but the pulsations
of the secondary need to be taken into account.

The effects of pulsations are most noticeable in the first two moments
of the absorption lines: shifts in the wavelength, and changes in line
width.  We therefore decided to allow the first two moments of the
secondary lines to vary freely for each observation obtained during a
primary or secondary eclipse. Each time a trial model is compared to
the data the position and width of the lines are adjusted. This scheme
is similar to the scheme for the normalization, but now focusing on
the lines of the secondary measured during eclipses. The average shift
in velocity is about $2$~km\,s$^{-1}$ and never larger than
$3$~km\,s$^{-1}$. The changes in width are always smaller then 2\%.

\begin{table*}
  \begin{center}
    \caption{Results for the CV\,Vel system.\label{tab:cvvel_results}}
    \smallskip 
    \begin{tabular}{l  r@{$\pm$}l r@{$\pm$}l   }
      \tableline\tableline
      \noalign{\smallskip}
      Parameter &  \multicolumn{2}{c}{ This work}  &    \multicolumn{2}{c}{Literature values} \\
      \noalign{\smallskip}   
      \hline
      \noalign{\smallskip}
      \multicolumn{5}{c}{Orbital parameters} \\
      \noalign{\smallskip}
      \hline
      \noalign{\smallskip}
      Time of primary minimum, T$_{\rm min, I}$ (BJD$-$2\,400\,000) & 42048.66944&0.00006 & 42048.66947&0.00014$^1$$^{\rm a}$ \\
      Period, $P$ (days)                             &  6.8894976&0.00000008 & 6.889494&0.000008$^1$ \\
      Cosine of orbital inclination, $\cos i_{o}$       &  0.0060&0.0003  &  \multicolumn{2}{c}{} \\   
      Orbital inclination, $i_{o}$ (deg)             & 86.54&0.02  & 86.59&0.05$^1$\\
      Velocity semi-amplitude primary, $K_{\rm p}$ (km\,s$^{-1}$)   & 126.69&0.035(stat)$\pm$0.1(sys) & 127.0&0.2$^2$ \\
      Velocity semi-amplitude secondary, $K_{\rm s}$ (km\,s$^{-1}$) & 129.15&0.035(stat)$\pm$0.1(sys) & 129.1&0.2$^2$ \\
      Velocity offset, $\gamma_{p}$ (km\,s$^{-1}$)   & \multicolumn{2}{c}{24.4$\pm$0.1 23.2$\pm$0.2$^{\rm b}$}& 23.9&0.3$^3$ \\
      Velocity offset, $\gamma_{s}$ (km\,s$^{-1}$)   & \multicolumn{2}{c}{24.6$\pm$0.1 23.3$\pm$0.2$^{\rm b}$}& 24.3&0.4$^3$ \\
      Orbital semi-major axis,    $a$ ($R_{\odot}$)  &  34.9&0.02 &  34.90&0.15$^2$  \\
      \noalign{\smallskip}
      \hline
      \noalign{\smallskip}
      \multicolumn{5}{c}{Stellar  parameters} \\
      \noalign{\smallskip}
      \hline 
      \noalign{\smallskip}
      Light ratio at 4480\,\AA{}, $L_{\rm s}/L_{\rm p}$ & 0.954&0.003 & 0.90&0.02$^1$ \\
      Fractional radius of primary, $r_{\rm p}$   & 0.1158&0.0002$^{\rm c}$  &  0.117&0.001$^1$ \\
      Fractional radius of secondary, $r_{\rm s}$ & 0.1139&0.0002$^{\rm c}$  &  0.113&0.001$^1$ \\
      $u_{1, \rm i} $+$u_{2, \rm i} $                                  & 0.35&0.1 & \multicolumn{2}{c}{(0.341+0.074)$\pm$0.1$^2$$^{\rm c}$} \\ 
      Macroturbulence broadening parameter,  $\zeta$ (km\,s$^{-1}$)  &  2.3&0.5$^{\rm d}$  & \multicolumn{2}{c}{} \\
      Microturbulence~$+$~PSF broadening parameter (km\,s$^{-1}$)             &  8.2&0.1$^{\rm d}$  & \multicolumn{2}{c}{} \\
      Projected rotation speed, primary, $v \sin i_{\rm p}$ (km\,s$^{-1}$)  & 21.5&0.3$\pm$2 & \multicolumn{2}{c}{See Table~\ref{tab:cvvel_vsini}} \\
      Projected rotation speed, secondary, $v \sin i_{\rm s}$ (km\,s$^{-1}$)& 21.1&0.2$\pm$2 & \multicolumn{2}{c}{See Table~\ref{tab:cvvel_vsini}} \\
      Projected spin-orbit angle, primary, $\beta_{\rm p}$ ($^{\circ}$)     & $-52.0$&$0.7$$\pm$6 & \multicolumn{2}{c}{} \\ 
      Projected spin-orbit angle, secondary, $\beta_{\rm s}$  ($^{\circ}$)  & $3.7$&$1.4$$\pm$7 & \multicolumn{2}{c}{} \\ 
      Primary mass,   $M_{\rm p} $ ($M_{\odot}$)   &  6.067&0.011$^{\rm d}$ & 6.066&0.074$^2$ \\
      Secondary mass, $M_{\rm s} $ ($M_{\odot}$)   &  5.952&0.011$^{\rm d}$ & 5.972&0.070$^2$ \\
      Primary radius,  $R_{\rm p} $ ($R_{\odot}$)  &  4.08&0.03$^{\rm e}$ & 4.126&0.024$^2$ \\
      Secondary radius, $R_{\rm s} $ ($R_{\odot}$) &  3.94&0.03$^{\rm e}$ & 3.908&0.027$^2$ \\     
      Primary $\log g_{\rm p}$ (cgs)               &  4.000&0.008 & 3.99&0.01$^1$ \\
      Secondary $\log g_{\rm s}$ (cgs)             &  4.021&0.008 & 4.03&0.01$^1$ \\
      \noalign{\smallskip}
      \tableline
      \noalign{\smallskip}
      \noalign{\smallskip}
      \multicolumn{5}{l}{{\sc Notes} ---}\\
      \multicolumn{5}{l}{$^{\rm a}$ We have placed the HJD$_{\rm UTC}$ value $2442048.66894$
        given by \cite{clausen1977} onto the BJD$_{\rm TDB}$ system.}\\
       \multicolumn{5}{l}{ \,\,\,\,The two systems differ by $46$ seconds
        ($0.00053$ days) for this particular epoch. }\\
      \multicolumn{5}{l}{$^{\rm b}$ The first value was calculated
        using the VALD line list; the second value was based on the rest frame wavelengths }\\
      \multicolumn{5}{l}{given by \cite{petrie1953}, provided here for
        continuity with previous works.}\\
      \multicolumn{5}{l}{$^{\rm c}$ Value was used as prior.}\\
      \multicolumn{5}{l}{$^{\rm d}$ See text for a discussion on the uncertainties}\\
      \multicolumn{5}{l}{$^{\rm e}$ Adopting a solar radius of $6.9566\cdot10^{8}$~m.}\\
      \noalign{\smallskip}
      \multicolumn{5}{l}{{\sc References} ---}\\
      \multicolumn{5}{l}{ (1) \cite{clausen1977} (2) \cite{yakut2007} (3) \cite{andersen1975}}
    \end{tabular}
  \end{center}
\end{table*}

Along with the spectroscopic data, we fitted the photometric data
described in Section~\ref{sec:data}. Because the eclipses last nearly
$12$~hr the data was obtained during different nights and cover large
ranges in airmass. We found that, even after performing differential
photometry on several comparison stars, the measured flux exhibits
significant trends with airmass. Therefore, for each nightly time
series, we added two parameters describing a linear function of
airmass which were optimized upon each calculation of $\chi^2$. As
mentioned in Section~\ref{sec:data} we also fitted the Str\"{o}mgren
{\it uvby} photometry from \cite{clausen1977}.
 
To constrain the quadratic limb darkening parameters $u_{1, \rm i}$
and $u_{2, \rm i}$ for the relevant bandpasses, we used the
'jktld'\footnote{{\tt
    http://www.astro.keele.ac.uk/jkt/codes/jktld.html}} tool to query
the predictions of ATLAS atmosphere models \citep{claret2000}. We
queried the models for the spectroscopic region (around $4500$~\AA{}),
the 'Ic' band, and the Str\"{o}mgren {\it uvby} observations by
\cite{clausen1977}. We placed a Gaussian prior on $u_1 + u_2$ with a
width of $0.1$ and held the difference $u_1 - u_2$ fixed at the
tabulated value. As the two known members of the CV\,Vel system are of
similar spectral type we used the same values for $\zeta$, $u_1$,
$u_2$, and the line strengths for both components.

Similar to the other groups who studied this system, we do not find
any sign of an eccentric orbit during our initial trials. We therefore
decided to set $e\equiv0$, in agreement with the results by
\cite{clausen1977}, who had gathered the most complete eclipse
photometry of the system to date. We found no sign of a systemic
drift in $\gamma$ over the three years of observations, and therefore
we did not include a linear drift term in our model. However, this
does not translate into a stringent constraint on the presence of a
potential third body, because most of our observations took place in
2010/2011. As described above, we used the line list from VALD in our
model. To derive results which can be compared with earlier works, we
also ran our model using the rest wavelengths given by
\cite{petrie1953}. For the obliquity work we prefer the VALD line
list, as it allows us to treat the \ion{Mg}{ii} as doublet, which is
important because of the relatively slow rotation in CV\,Vel
(Figure~\ref{fig:cvvel_lines}).  Table~\ref{tab:cvvel_results}
presents the $\gamma$ values from both runs.\footnote{\cite{yakut2007}
  used different wavelengths for \ion{Mg}{ii} which lead to a
  different values of $\gamma_{\rm p}$ and $\gamma_{\rm s}$. Adjusting
  for the difference in the wavelength position we find that the
  results by \cite{yakut2007} are also consistent with the results by
  \cite{andersen1975}.}

\begin{figure*}
  \begin{center}
    \includegraphics[width=8.6cm]{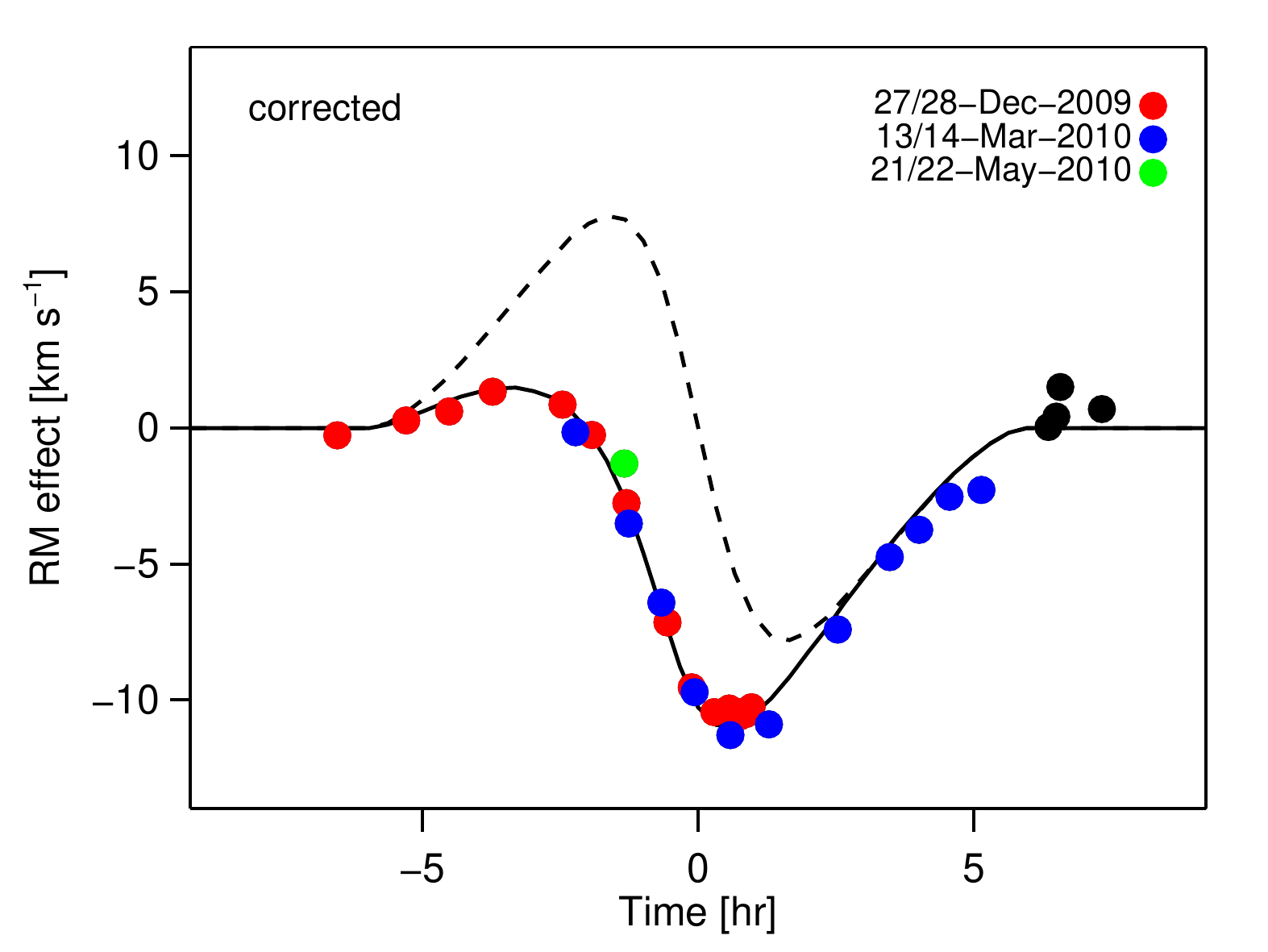} 
    \includegraphics[width=8.6cm]{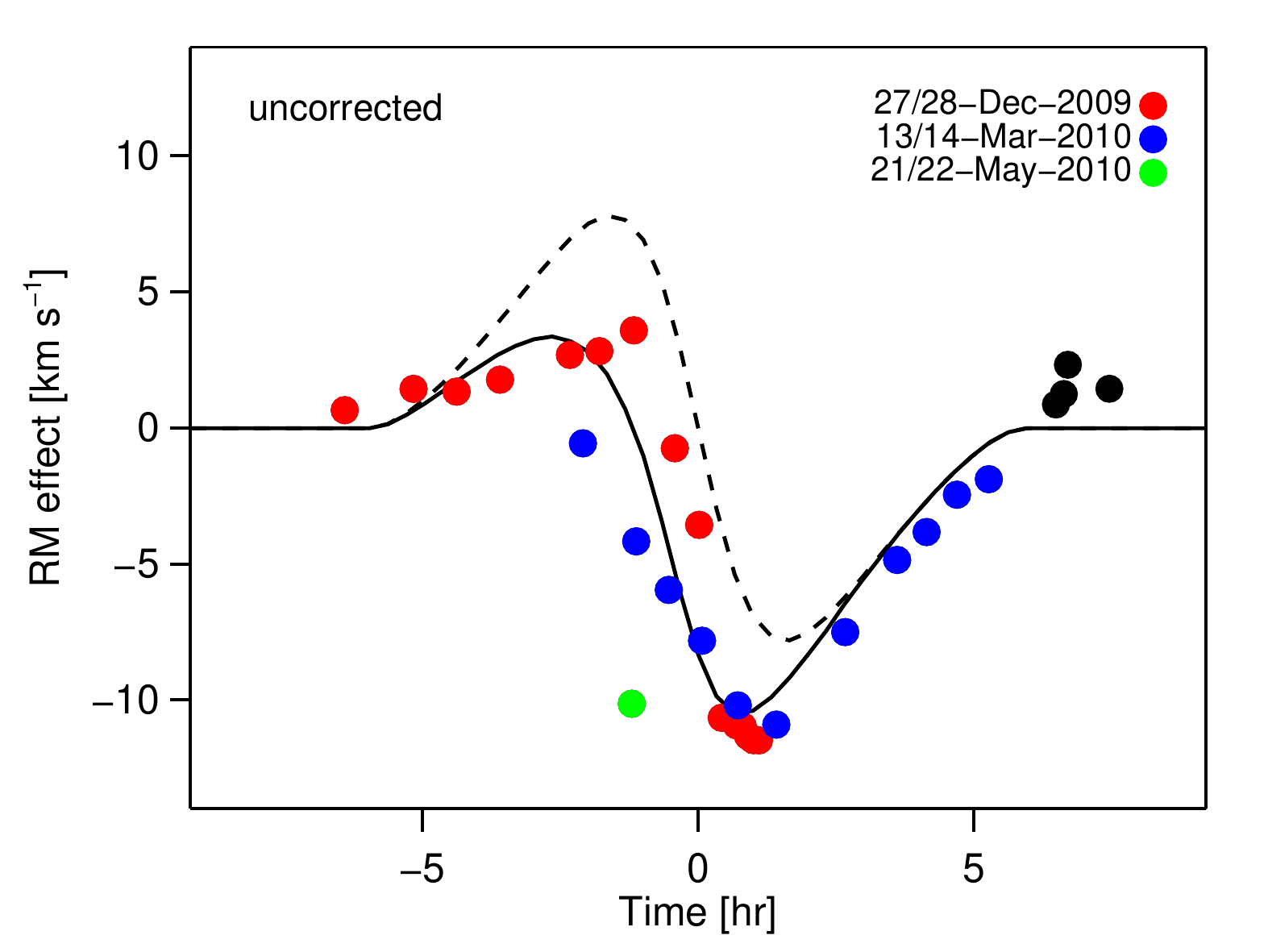}    
    \caption{\label{fig:cvvel_primary_RM} {\bf Anomalous RVs during
        primary eclipses of CV\,Vel.} {\it Left.}---Apparent RV
      of the primary, after subtraction of the best-fitting orbital
      model. The solid line is the best-fitting model for the RM effect.
      The dashed line indicates the expected signal for a well-aligned system.
      {\it Right.}---Same, but neglecting any correction for pulsations
      of the secondary.}
  \end{center}
\end{figure*}

\begin{figure}
  \begin{center}
    \includegraphics[width=8.6cm]{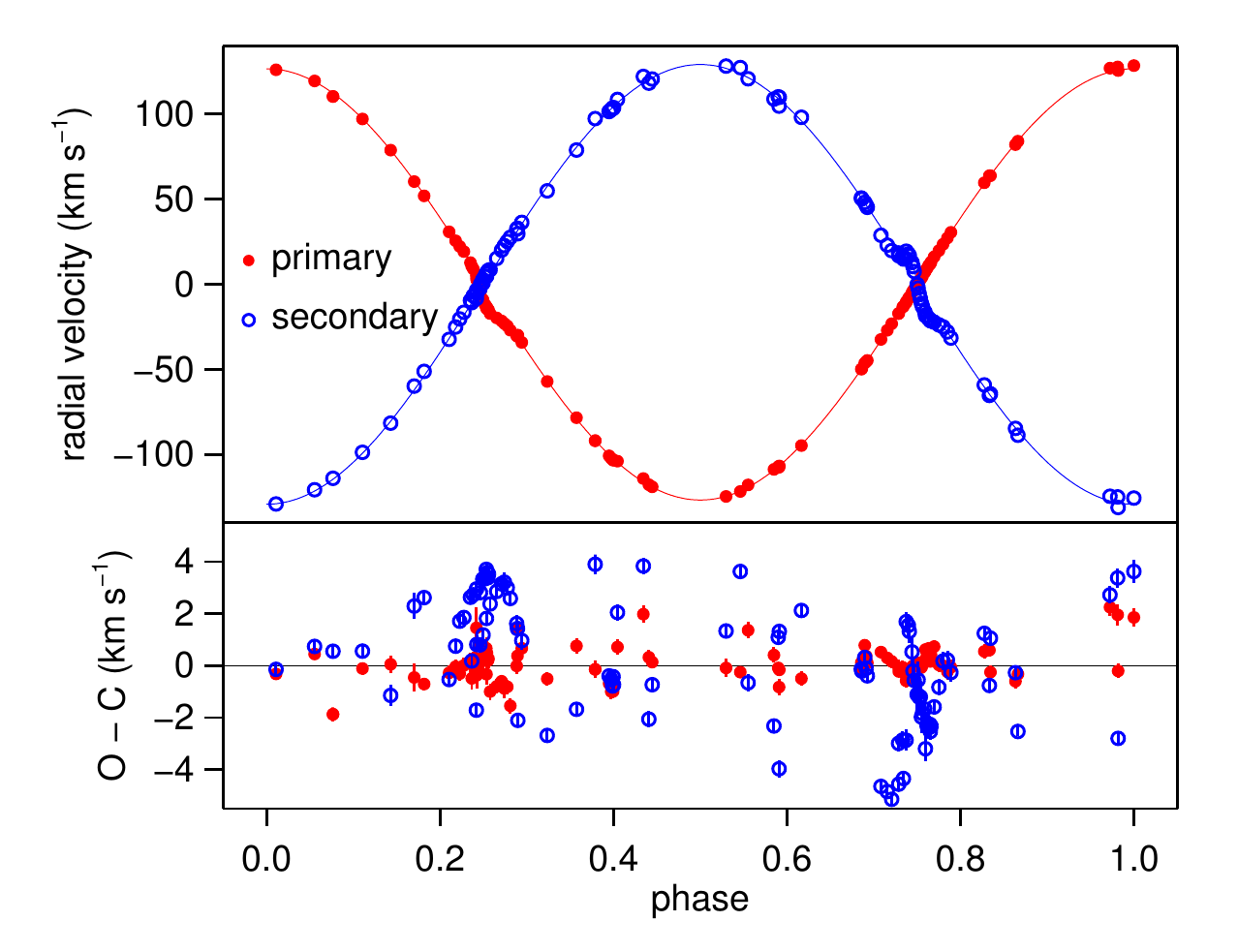} 
       \caption{\label{fig:cvvel_orbit} {\bf Orbit of CV\,Vel} {\it
        Top.}---Apparent RVs for the primary and secondary stars in
      the CV\,Vel system. This plot is similar to
      Figure~\ref{fig:cvvel_primary_RM} but now for all RVs taken
      2009-2013. These RVs are shown for illustrative purposes only
      and are not used in the analysis of the system. 
      {\it Bottom.}--- RVs minus best fitting model from the fit to
      the line shapes. The secondary star exhibits, due to its larger
      pulsations (Figure~\ref{fig:cvvel_lines}), a larger RV scatter
      than the primary star.}
  \end{center}
\end{figure}

\subsection{Results}
\label{sec:cvvel_results}

The results for the model parameters are given in
Table~\ref{tab:cvvel_results}.
Figure~\ref{fig:cvvel_primary_eclipse_grayscale} shows a grayscale
representation of the primary spectra in the vicinity of the
\ion{Mg}{ii} line during the eclipse.
Figures~\ref{fig:cvvel_secondary_eclipse_grayscale} shows the same for
the spectra obtained during secondary
eclipses. Figure~\ref{fig:cvvel_primary_eclipse_lines} presents a
  subset of primary eclipse observations in a more traditional way.

Concerning the orbital parameters, we find results that are consistent
with earlier works. The uncertainties in the fractional stellar radii
are small, with significant leverage coming from the spectroscopic
eclipse data.  Since we have not fully explored how the pulsations in
the absorption lines influence our results for the scaled radii, for
the purpose of calculating absolute radii we have taken the
conservative approach of using the previously determined fractional
radii, which have larger uncertainties from
\cite[][$r_p=0.117\pm0.001$ and $r_p=0.113\pm0.001$]{clausen1977}. We
have not tested how the exact timing of the observations in
combination with the pulsations might influence the results for the
velocity semi-amplitudes and suggest that $0.1$~km\,s$^{-1}$ is a more
realistic uncertainty interval for $K_{\rm i}$ then the statistical
uncertainty of  $0.035$~km\,s$^{-1}$. We use the enlarged uncertainties
in calculating the absolute dimensions of the system. The results for the
macroturbulent width $\zeta$ and the microturbulent/PSF width are
strongly correlated. The inferred breakdown between these types of
broadening depends on our choice of limb darkening parameters. At this
point we can only say that any additional broadening beyond rotation
is about $8-9$~km\,s$^{-1}$.

\begin{figure}
  \begin{center}
    \includegraphics[width=8.6cm]{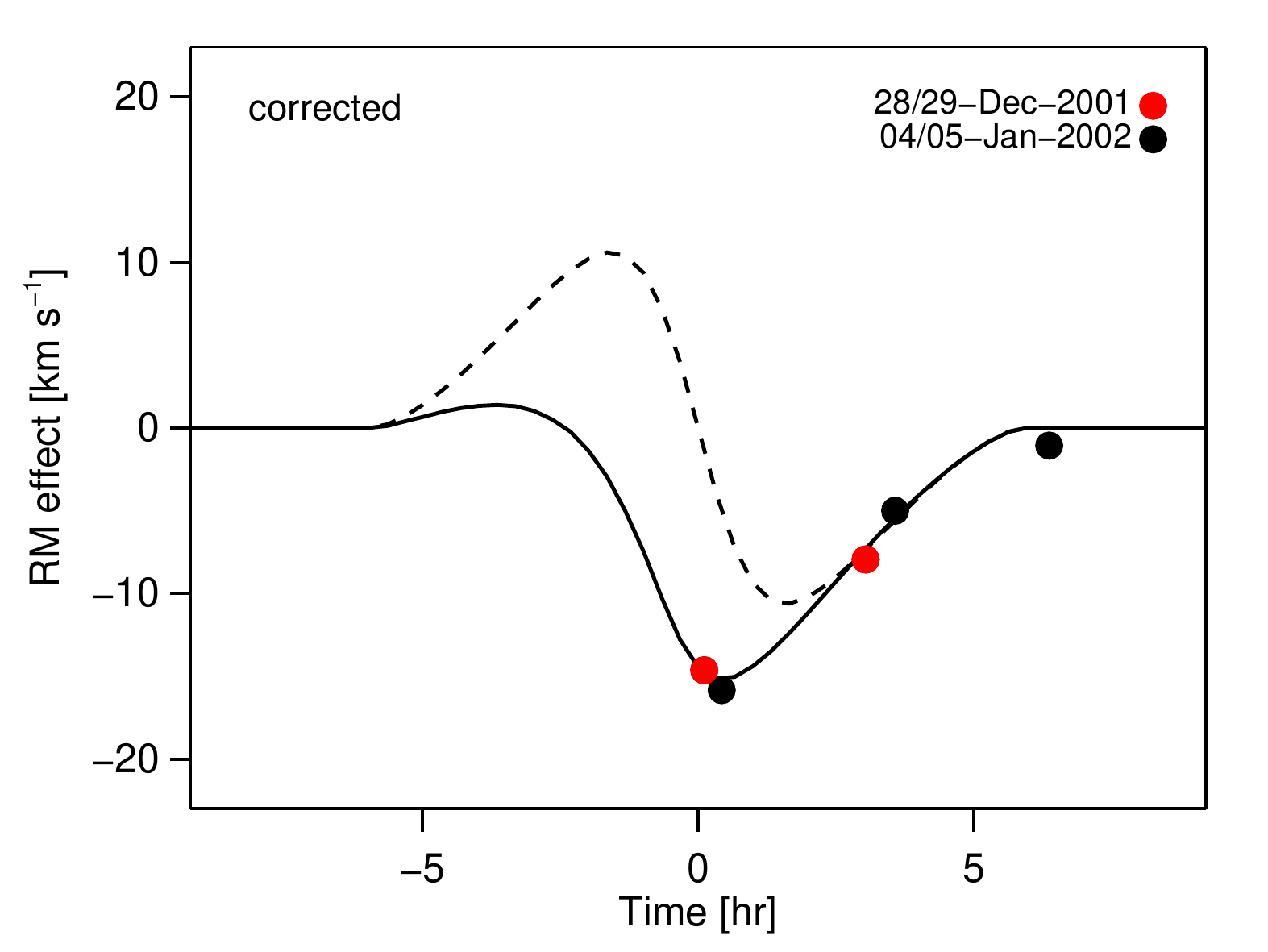} 
    \caption {\label{fig:cvvel_primary_RM_yakut} {\bf Anomalous RVs
        during primary eclipses 2001/2002 in the CV\,Vel system.}
      Same as Figure~\ref{fig:cvvel_primary_RM}, but for the data from
      2001/2002 obtained by \cite{yakut2007}.  Note the larger scale
      of the $y$-axis compared to the two panels in
      Figure~\ref{fig:cvvel_primary_RM}. This is because the $v \sin
      i_\star$ of the primary was higher during 2001/2002.}
  \end{center}
\end{figure}

\paragraph{Projected obliquities and projected rotation speeds} We
find that the sky projection of the primary rotation axis is
misaligned against the orbital angular momentum, with $\beta_{\rm
  p}=-52.0\pm0.7^{\circ}$. The projection of the secondary axis appears
to be aligned ($\beta_{\rm s}=3.7\pm1.4^{\circ}$).

Our method of correcting for pulsations of the secondary turned out to
be important, but even with no such corrections the result of a
misaligned primary is robust. This is illustrated in
Figure~\ref{fig:cvvel_primary_RM}, which shows the anomalous RVs
during primary eclipse. To create this figure we subtracted our
best-fitting model of the secondary spectrum from each of the observed
spectra.  We then measured the RV of the primary star at each epoch,
by fitting a Gaussian function to the \ion{Mg}{ii} line.  We then
isolated the RM effect by subtracting the orbital RV, taken from the
best-fitting orbital model.  The right panel in
Figure~\ref{fig:cvvel_primary_RM} shows the results for the case when
no correction was made for pulsations. There is evidently scatter
between the results from different nights, but the predominance of the
blueshift throughout the transit implies a misaligned system (a formal
fit gives $\beta_{\rm p}=-37^{\circ}$ and $\beta_{\rm s}=-1^{\circ}$).
The left panel shows the results for the case in which we have
corrected for the time variations in the first two moments of the
secondary lines. The scatter is much reduced and the fit to the
geometric model is much improved.

Figure~\ref{fig:cvvel_orbit} shows for completeness all RVs
obtained. As mentioned above the RVs out of eclipse have not been
corrected for the influence of pulsations. None of the RVs are used in
the analysis they are shown here for comparison only.

We further repeated our analysis on two additional lines, the
\ion{Si}{iii} line at $4552.6$~\AA{} and the \ion{He}{I} line at
$6678$~\AA{}. For \ion{Si}{iii} we obtain $\beta_{\rm p}=-58^{\circ}$
and $\beta_{\rm s}=-4^{\circ}$, and for \ion{He}{I} we measure
$\beta_{\rm p}=-52^{\circ}$ and $\beta_{\rm s}=-1^{\circ}$. The
\ion{Si}{iii} line is weaker than the \ion{Mg}{ii} line and the
\ion{He}{I} is pressure broadened, which make the analysis more
complex \citep{albrecht2011}. We therefore prefer the result from the
\ion{Mg}{ii} line. However we judge that the total spread in the
results $6^\circ$ and $7^\circ$ are probably closer to the true
uncertainty in the projected obliquities, than our formal errors. This
is because our formal uncertainty intervals rely on measurements taken
during 3 and 4 nights, for the primary and secondary,
respectively. For a better uncertainty estimation measurements
obtained during more different nights, or a more carefully handling of
the pulsations, would be needed.

\begin{table}
  \begin{center}
    \caption{Projected rotation speed measurements in the in CV\,Vel
      system.}
    \label{tab:cvvel_vsini}
    \begin{tabular}{l r@{$\pm$}l r@{$\pm$}l r@{$\pm$}l }
      \hline
      \hline      
      \noalign{\smallskip}
      Year & \multicolumn{2}{l}{$1973$}                       & \multicolumn{2}{l}{$2001-2002$}      & \multicolumn{2}{l}{$2009-2010$}\\
      Ref.      & \multicolumn{2}{l}{\cite{andersen1975}} & \multicolumn{2}{l}{\cite{yakut2007}}  & \multicolumn{2}{l}{This work}\\
      \noalign{\smallskip}
      \hline
      \noalign{\smallskip}
      $v \sin i_{\rm p}$ (km\,s$^{-1}$)   &  28&3  &  29.5&$2^{\rm a}$ & 21.5&2  \\
      $v \sin i_{\rm s}$ (km\,s$^{-1}$)   &  28&3  &  19.0&$2^{\rm a}$ & 21.1&2  \\
      \noalign{\smallskip}
      \hline
      \tableline
      \noalign{\smallskip}
      \noalign{\smallskip}
      \multicolumn{7}{l}{{\sc Notes} ---}\\
      \multicolumn{7}{l}{$^{\rm a}$Based on our own analysis of the
        \cite{yakut2007} spectra. }\\
    \end{tabular}
  \end{center} 
\end{table}

For the projected rotation speeds we find $v \sin i_{\rm
  p}=21.5\pm0.3$~km\,s$^{-1}$ and $v \sin i_{\rm
  s}=21.1\pm0.2$~km\,s$^{-1}$. Making the same measurement in the
\ion{Si}{iii} lines one would obtain $v \sin i_{\rm p}=
20.6$~km\,s$^{-1}$ and $v \sin i_{\rm s}= 20.0$~km\,s$^{-1}$. For
similar reasons as mentioned above for the projected obliquity we
suspect that also the formal uncertainties for $v \sin i_\star$ are
underestimated. In what follows we assume that an uncertainty of
$2$~km\,s$^{-1}$ is appropriate.

\cite{yakut2007} obtained $4$ of their $30$ observations during
primary eclipses. We performed a similar analysis of their spectra, in
the same manner as our own data.  For the projected obliquity of the
primary star in 2001/2002 we obtained $\beta_{\rm p\,
  2002}=-55\pm3^{\circ}$. As this result rests mainly on two
observations, obtained nearly at the same eclipse phase
(Figure~\ref{fig:cvvel_primary_RM_yakut}), we judge that the true
uncertainty is much larger, probably about $15^\circ$. For the
projected rotation speeds we obtained $v \sin i_{\rm p\,
  2002}=29.5\pm2$~km\,s$^{-1}$ and $v \sin i_{\rm s\,
  2002}=19.0\pm2$~km\,s$^{-1}$, adopting a conservative uncertainty
interval as we did for our spectra.

\begin{figure}
  \begin{center}
    \includegraphics[width=8.5cm]{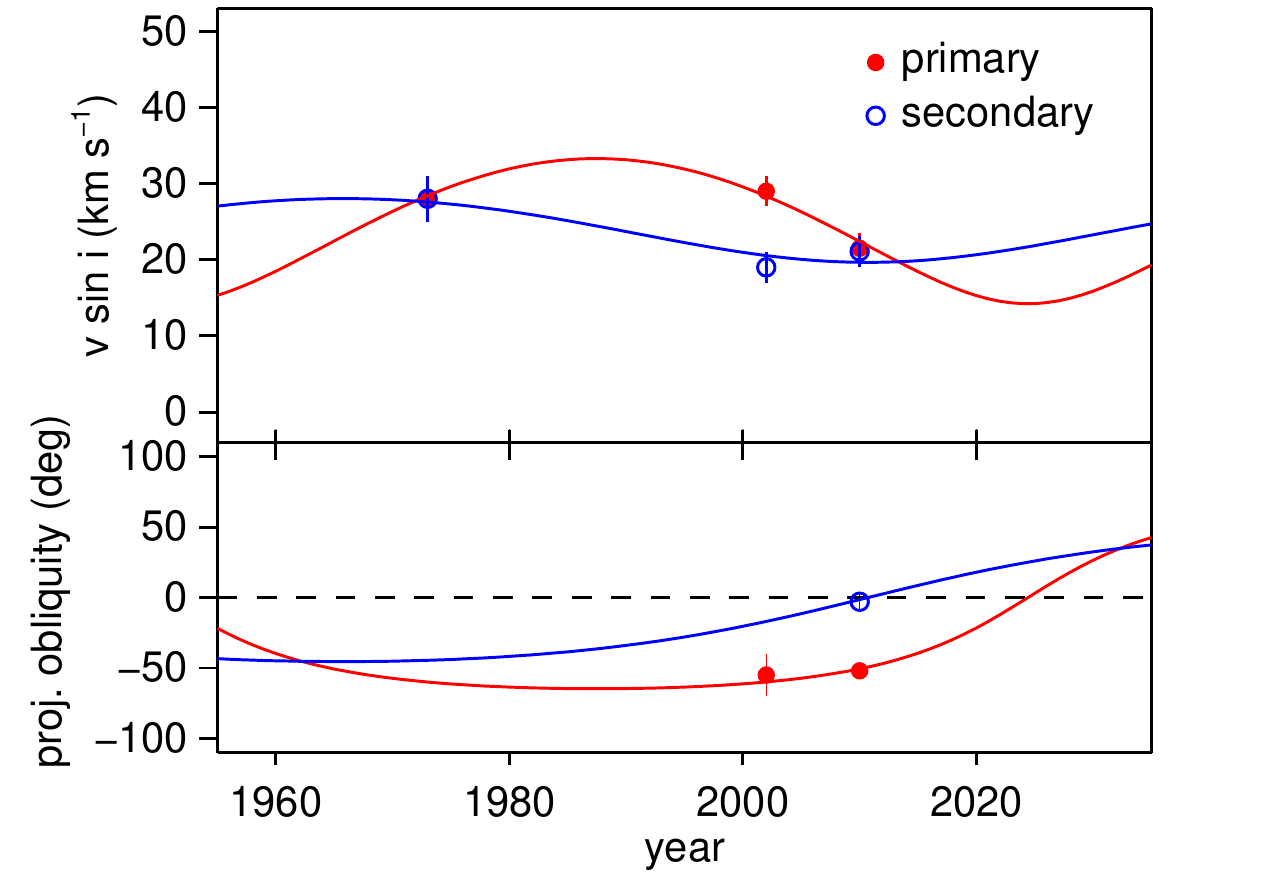} 
    \caption  {\label{fig:cvvel_axes} {\bf Precession of the stellar
        rotation axes in the CV\,Vel system.} The upper panel shows the
      measured $v \sin i_\star$ values of the primary (red solid
      symbol) and secondary (blue open symbol) stars. The lines
      indicate the time evolution in our best-fitting
      model. The lower panel shows the measurements of $\beta$ and the
      predictions of our model.}
  \end{center}
\end{figure}

\section{Rotation and Obliquity}
\label{sec:cvvel_rot}

\subsection{Precession of the Rotation Axes}

Our results for $v \sin i_{\star}$ differ from the values found by
\cite{andersen1975} and from the values found by \cite{yakut2007} (see
Table~\ref{tab:cvvel_vsini}). Evidently the projected rotation rates
are changing on a timescale of decades. We are only aware of a few
cases in which such changes have been definitively observed, one being
the DI\,Her system \citep{reisenberger1989,albrecht2009,philippov2013}.

\cite{yakut2007} used the {\it CORALIE} spectrograph on the $1.2$~m
Swiss telescope for their observations. To exclude any systematic
effects due to the choice of instrument---as unlikely as it might
seem---we also collected a number of spectra with the {\it CORALIE}
spectrograph, as described in Section~\ref{sec:data}, which confirmed
the time variation of the $v\sin i_{\star}$ of the primary. The line
width of the secondary appears to have changed between the
observations conducted by \cite{andersen1975} and \cite{yakut2007}.

Assuming that $v$ remained constant over the interval of observations
($\sim30$~yr), we interpret these results as variations in $\sin
i_{\star}$ for both stars. This allows us to learn about the
precession rates of the stellar rotation axes around the total angular
momentum vector of the system. Employing the formulas from
\cite{reisenberger1989} we can use the $v \sin i_{\star}$ values from
Table~\ref{tab:cvvel_vsini} together with our measurements of the
projected obliquity to obtain values for the stellar obliquities
($\psi$) and rotation velocities of the two stars.

For this purpose, in addition to the system parameters of CV\,Vel
which are presented in Table~\ref{tab:cvvel_vsini}, we need values for
the apsidal motion constant ($k_2$) and the radius of gyration
($\theta$) of each star. These we obtain from the tables presented by
\cite{claret2004b}. We use the model with a mass of $6.3$~M$_{\odot}$,
close to the mass of the stars in the CV\,Vel system, and estimate the
uncertainty by considering the age interval from $30-50$~Myr. The age
of CV\,Vel is estimated to be $40$~Myr \citep{yakut2007}. The results
for both stars are $k_2=0.005\pm0.002$ and $\theta=0.044\pm0.012$.

Using these values we carried out a Monte Carlo experiment, in which
we draw system parameters by taking the best-fitting values and adding
random Gaussian perturbations with a standard deviation equal to the
1$\sigma$ uncertainty. For each draw, we minimize a $\chi^2$ function
by adjusting $\psi$ and $v$ for each star, as well as the particular
times when the spin and orbital axes are aligned on the
sky. Furthermore we allow the $k_2$ and $\theta$ values to vary with a
penalty function given by the prior information mentioned above. The
resulting parameters are presented in
Table~\ref{tab:cvvel_precession}. In Figure~\ref{fig:cvvel_axes} we
show the data for $v \sin i_{\star}$ and $\beta$, as well as our model
for their time evolution. We obtain $\psi_{\rm p}= 67\pm4^{\circ}$ and
$\psi_{\rm s}= 46\pm9^{\circ}$ for the obliquities, and $v_{\rm p}=
35\pm5$~km\,s$^{-1}$ and $v_{\rm s}= 28\pm4$~km\,s$^{-1}$ for the
rotation speeds.

The formal uncertainties for $\psi$ and $v$ should be taken with a
grain of salt. We did not observe even half a precession period, which
makes an estimation of $\psi$ and $v$ strongly dependent on our
assumptions regarding $k_2$ and $\theta$. We have only a small number
of measurements: $3$ $v \sin i_{\star}$ and one or two $\beta$
measurement per star, amounting to $9$ data points. With these we aim
to constrain $6$ parameters: $v$, $\psi$, and a reference time for
each star. In this situation we can determine parameter values, but we
cannot critically test our underlying assumptions. For the secondary
in particular we have only little information to constrain $\psi$ and
$v$. The only indication we have for this star that it is not aligned
is the change in $v \sin i_{\star}$ between 1973 and
2001/2002. Clearly, future observations would be helpful to confirm
the time variations. Measurements of the projected obliquity in only a
few years should be able to establish if this star's axis is indeed
misaligned (Figure~\ref{fig:cvvel_axes}). Finally we obtain somewhat
different values for $k_2$ and $\theta$ for the two stars, which have
similar masses and the same age. This is because for the primary the
fast change in $v \sin i_\star$ between 2002 and 2010 requires a fast
precession timescale.

\begin{table}
  \caption{Precession of the stellar axes in CV\,Vel}
  \label{tab:cvvel_precession}
  \begin{center}
    \smallskip
    \begin{tabular}{l r@{$\pm$}l }
      \hline
      \hline
      \noalign{\smallskip}
      Parameter &\multicolumn{2}{c}{CV\,Vel}  \\
      \noalign{\smallskip}
      \hline
      \noalign{\smallskip}
      Rotation speed of primary $v_p$  (km\,s$^{-1}$) &  33&4     \\
      Rotation speed of secondary $v_S$ (km\,s$^{-1}$)  &  28&4     \\
      Obliquity of primary $\psi_{\rm p}$ ($^{\circ}$)  & $64$&4  \\ 
      Obliquity of secondary $\psi_{\rm s}$ ($^{\circ}$)  &  $46$&9  \\ 
      Year when $\beta_{\rm p} = 0^\circ$ & 2023&7   \\
      Year when $\beta_{\rm s} = 0^\circ$ &  2011&4   \\
      Radius of gyration of primary  $\theta_{\rm p}$  &  0.0363&0.0095  \\ 
      Radius of gyration of secondary  $\theta_{\rm s}$  &  0.0451&0.0009  \\ 
      Apsidal motion constant of primary $k_{2,{\rm p}}$  &  0.0063&0.0007  \\
      Apsidal motion constant of  secondary $k_{2,{\rm s}}$  & 0.0047&0.0003    \\
      \noalign{\smallskip}
      \hline
      \noalign{\smallskip}
      Precession period of primary (yr) &  139&54$^\star$  \\
      Precession period of secondary (yr) & 177&22$^\star$ \\
      \noalign{\smallskip}
      \noalign{\smallskip}
      \hline
      \noalign{\smallskip}
      \noalign{\smallskip}
    \end{tabular}
    \tablecomments{ 
      $^\star$ The precession periods are not free parameters in this
      model. They are derived from the obliquity, the internal
      structure parameters, and the rotation speeds.} 
  \end{center} 
\end{table}

The last point could reflect a shortcoming of our simple model (some
missing physics), an underestimation of the errors in the $v \sin
i_\star$ measurements or the presence of a third body.  Nevertheless
the finding of a large projected misalignment for the primary and the
changes in $v \sin i_{\star}$ measured for both stars makes it
difficult to escape the conclusion that the stars have a large
obliquity and precess, even if the precise values are difficult to
determine at this point. A more detailed precession model and more
data on $\beta$ and $v \sin i_{\star}$, obtained over the next few
years, would help in drawing a more complete picture.

We note that in principle, one can also use the effect of gravity
darkening on the eclipse profiles to constrain $\psi$, as was done
recently by \cite{szabo2011}, \cite{barnes2011} and
\cite{philippov2013} for the KOI-13 and DI\,Her systems. However as
the rotation speed in CV\,Vel is a factor few slower then in these two
systems this would require very precise photometric data. We also note
that small changes in the orbital inclination of CV\,Vel are expected,
as another consequence of precession. This might be detected with
precise photometry obtained over many years.

\subsection{Time evolution of the spins}
\label{sec:cvvel_evolution}

With an age of $40$~Myr \citep{yakut2007} CV\,Vel is an order of
magnitude older than the even-more misaligned system DI\,Her
($4.5\pm2.5$~Myr, $\beta_p=72\pm4^{\circ}$ $\beta_s=-84\pm8^{\circ}$,
\citealt{albrecht2009,claret2010}).  In this section we investigate if
CV\,Vel could have evolved from a DI\,Her-like configuration, through
the steady action of tidal dissipation.  If so, CV\,Vel might
represent a link between young systems with large misalignment, and
older systems where tidal interactions have had enough time to attain
the equilibrium condition of a circular orbit with aligned and
synchronized spins.

In Paper~IV, we found that the EP\,Cru system (age $57\pm5$~Myr) could
not have evolved out of a DI\,Her like system, despite the strong
similarities of all the system parameters except the stellar obliquity
and age. This is because the $v \sin i_\star$ values in EP\,Cru are
about $9$ times the expected value for the pseudosynchronized state.
Theories of tidal interactions predict that damping of any significant
spin-orbit misalignment should occur on a similar same time scale as
synchronization of the rotation \citep{hut1981,eggleton2001}. This is
because in these tidal models, a single coefficient describes the
coupling between tides and rotation. If the stellar rotation frequency
is much larger than the synchronized value, then rotation around any
axis is damped at about the same rate.\footnote{\cite{lai2012}
recently suggested that, for the case of stars with an connective
envelope  -- stars of much lower mass then the stars we study here --
dynamical tides can damp different components of the stellar spin on very
different timescales.} Therefore while the rotation speed is reduced,
the angle between the overall angular momentum and stellar rotation
axis does not change. When the stellar rotation around the axis
parallel to the orbital angular momentum approaches the synchronized
value, then rotation around this axis becomes weakly coupled to the
orbit. Tidal damping of rotation around any other axis will only cease
when the rotation around these axes stops, and the stellar spin is
aligned with the orbital axis. Therefore, finding a system in an
aligned state that is rotating significantly faster than synchronized
rotation indicates, according to these tidal theories, that the
alignment was primordial. In Paper III, we found that NY\,Cep is also
inconsistent with having evolved from a state with large misalignment.

\begin{figure}
  \begin{center}
    \includegraphics[width=8cm]{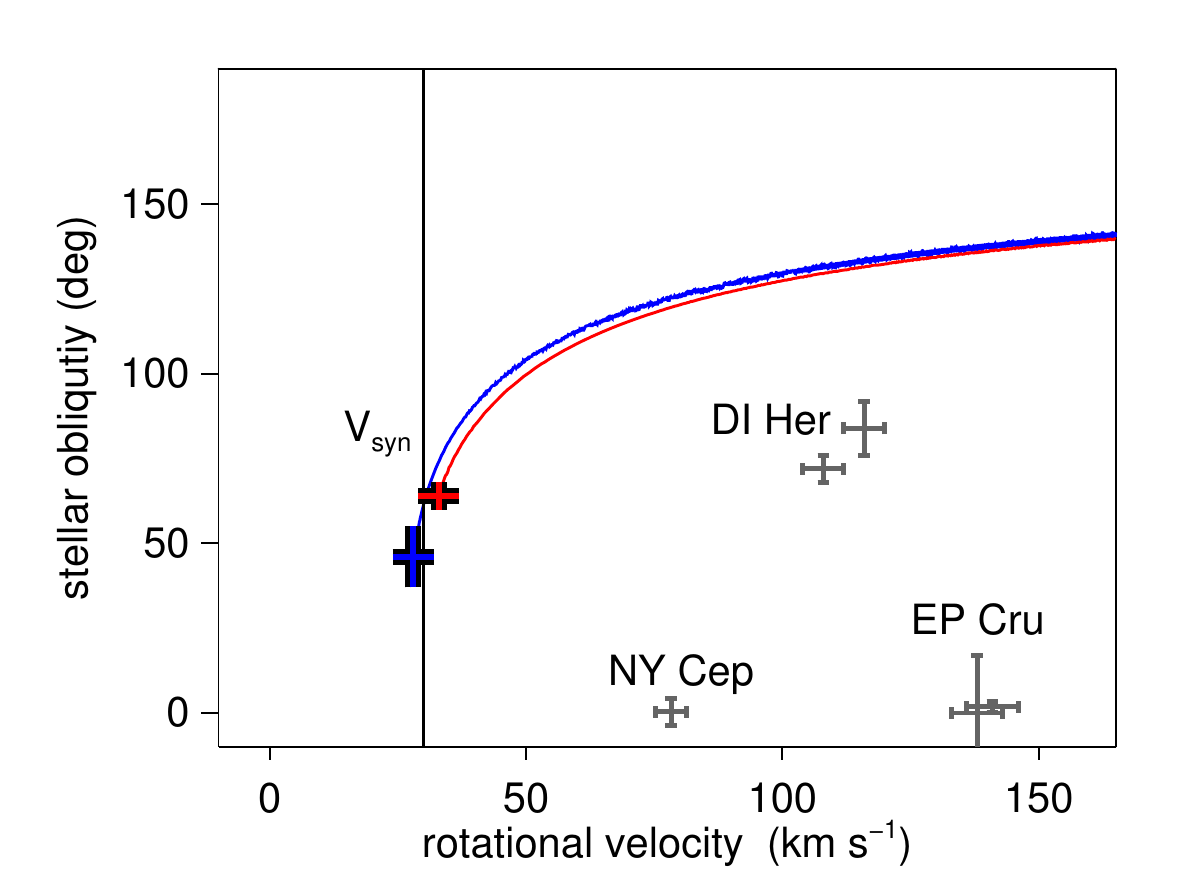} 
    \caption {\label{fig:cvvel_evolution} {\bf Tidal evolution of
        CV\,Vel.} The red and blue crosses mark the derived rotation
      speeds and obliquities of the primary and secondary.  Red and
      blue lines show the theoretical obliquity evolution of a system
      like CV\,Vel. Here we used the currently measured values for
      CV\,Vel and evolved the system back in time. The model includes
      the evolutionary changes in stellar radius with time, and adopts
      a viscous timescale ($t_{\rm V}$) of 300\,000~yr, about 6\,000
      times larger than what is normally assumed for late type
      stars. A lower value of $t_{\rm V}$ would lead to an overall
      faster tidal evolution. It will leave the ratio of the alignment
      and synchronization timescales unchanged.  According to these
      simulations and measurements it is conceivable that CV\,Vel had
      larger (DI\,Her-like) misalignments when it was younger, and is
      currently undergoing tidal realignment. The vertical line
      indicates the synchronized rotation speed ($V_{\rm syn}$) for
      the current orbital configuration and stellar radii of
      CV\,Vel. We also show the measured $v \sin i_\star$ and
      $\beta_\star$ for DI\,Her, NY\,Cep, and EP\,Cru, three other
      systems from the BANANA survey. While the  exact value of
      $V_{\rm syn}$ for these systems differ from the value for
      CV\,Vel,  all these systems do rotate significantly faster than
      their synchronized or pseudosynchronized values.}
  \end{center}
\end{figure}

For CV\,Vel, synchronized rotation would correspond to $v\approx
30$~km\,s$^{-1}$ for both stars. The slow rotation speeds and
misaligned axes suggest that we are observing this system in a state
in which tides are currently aligning the axes. To illustrate we use
the TOPPLE tidal-evolution code \citep{eggleton2001} with the
parameters from Table~\ref{tab:cvvel_results} and
Table~\ref{tab:cvvel_precession} and evolve the system backwards in
time. The results are shown in Figure~\ref{fig:cvvel_evolution}. It
appears as if the current rotational state of the two stars is
consistent with an evolution out of a higher-obliquity state. We
reiterate, though, that the rotation speed and obliquity of the lower
mass star are rather uncertain. The results of the tidal evolution do
depend on the exact parameters we use for CV\,Vel, taken from the
confidence intervals of our measurements. Therefore we can not make
strong statements about the exact evolution CV\,Vel has taken. However
the qualitative character of the evolution did remain the same in all
of our runs.

CV\,Vel did evolve out of a state with larger obliquities and faster
rotation. Under this scenario, we are seeing the system after only
about one obliquity-damping timescale, which implies that only a small
fraction of an eccentricity damping timescale has elapsed (due to the
angular momentum in the orbit being greater than that in the spins).

This appears to be a counterintuitive result as one would expect that
whatever creates high obliquities would also create a high
eccentricity, which should still be present, according to our simple
simulation. Scenarios involving a third body, may account for the
misaligned spins despite tidal damping of the eccentricity. For
example \cite{eggleton2001} showed that in the triple system SS\,Lac,
with inner and outer orbits non-parallel, the spin orientations of the
two inner components could vary on a timescale of just several hundred
years.

As long as the inner and outer orbits remain non-coplanar, the inner
orbit will precess around the total angular momentum. The orbital
precession timescale will most likely be not the same as the
precession timescale of the two stars. Therefore the angle between the
stellar spins and the orbital plane of the inner orbit can remain
large even after many obliquity damping timescales. The system would
settle into a Cassini state, with the oblique spins precessing at the
same rate as the inner orbit. A pseudo-synchronous spin rate would
settle in for the oblique yet circular orbits
\citep[e.g.][]{levrard2007,fabrycky2007b}. Of course such a scenario
remains speculative as long as no third body is searched for and
detected.

The state of the obliquities suggests that DI\,Her and CV\,Vel have a
history which is qualitatively different from the history of NY\,Cep,
and EP\,Cru. The two later systems had good alignment throughout their
main sequence lifetime, while DI\,Her and CV\,Vel did at some point
acquire a larger misalignment.

\section{Summary}
\label{sec:summary}

We have analyzed spectra and photometry of the CV\,Vel system,
obtained during primary and secondary eclipses as well as outside of
eclipses. Taking advantage of the Rossiter-McLaughlin effect, we find
that the rotation axis of the primary star is tilted by
$-52\pm6^{\circ}$ against the orbital angular momentum, as seen on the
sky. The sky projections of the secondary rotation axis and the
orbital axis are well aligned ($3\pm7^{\circ}$). Furthermore we find
that the projected rotation speeds ($v \sin i_\star$) of both stars
are changing on a timescale of decades. We interpret these changes as
a sign of precession of the stellar rotation axes around the total
angular momentum of the system. Using the $v \sin i_\star$
measurements (ours and literature measurements dating $30$ years back)
in combination with our projected obliquity measurements, we calculate
the rotation speed ($v$) as well as the true obliquity ($\psi$) of
both stars. We find obliquities of $\psi_{\rm p}= 64\pm4^{\circ}$ and
$\psi_{\rm s}= 46\pm9^{\circ}$ and rotation speeds of $v_{\rm p}=
33\pm4$~km\,s$^{-1}$ and $v_{\rm s}= 28\pm4$~km\,s$^{-1}$ for the two
stars. While the results for the primary star are relatively solid,
the results for the secondary star rely on changes in the measured
line width only, and need to be confirmed with future spectroscopic
observations.

Our results for the stellar rotation in CV\,Vel are consistent with
long-term tidal evolution from a state in which the stars had higher
rotation speeds as well as higher obliquities, similar to what we
found in the younger binary system DI\,Her. In this sense it seems
plausible that DI\,Her and CV\,Vel are two points on an evolutionary
sequence from misaligned to aligned systems. Given the simplest tidal
theories, the other systems in our sample (NY\,Cep, and EP\,Cru) could
not have realigned via tides. So far it is not clear what causes the
difference between these two groups. Given recent findings that close
binaries are often accompanied by a third body, it is tempting to
hypothesize that the influence of a third body is the key factor that
is associated with a large misalignment. No third body has yet been
detected in either the CV\,Vel nor DI\,Her systems, nor have these
systems been thoroughly searched.\footnote{ \cite{kozyreva2009} found
  a possible pattern in the eclipse timing of DI\,Her, indicating a
  third body. However \cite{claret2010} found no evidence for a third
  body, employing a dataset which includes the timings from
  \cite{kozyreva2009}.}  Such a search should be a priority for future
work.

\acknowledgments We would like to thank the anonymous referee for
timely suggestions, which improved the manuscript.  We thank Kadri
Yakut and Conny Aerts for providing us with a digital version of the
\cite{clausen1977} photometry, as well as their own {\it CORALIE}
spectra and comments on the manuscript. S.A.\ acknowledges support
during part of this project by a Rubicon fellowship from the
Netherlands Organisation for Scientific Research (NWO). Work by S.A.\
and J.N.W.\ was supported by NASA Origins award NNX09AB33G and NSF
grant no.\ 1108595. TRAPPIST is a project funded by the Belgian Fund
for Scientific Research (FNRS) with the participation of the Swiss
National Science Fundation (SNF).  MG and EJ are FNRS Research
Associates. A. H.M.J.\ Triaud received funding from of a fellowship
provided by the Swiss National Science Foundation under grant number
PBGEP2-14559. This research has made use of the following web
resources: {\tt simbad.u-strasbg.fr, adswww.harvard.edu,arxiv.org,
  http://arxiv.org}


\begin{thebibliography}{45}
\expandafter\ifx\csname natexlab\endcsname\relax\def\natexlab#1{#1}\fi

\bibitem[{{Albrecht} {et~al.}(2009){Albrecht}, {Reffert}, {Snellen}, \&
  {Winn}}]{albrecht2009}
{Albrecht}, S., {Reffert}, S., {Snellen}, I.~A.~G., \& {Winn}, J.~N. 2009,
  \href{http://dx.doi.org/10.1038/nature08408}{\nat, 461, 373}

\bibitem[{{Albrecht} {et~al.}(2007){Albrecht}, {Reffert}, {Snellen},
  {Quirrenbach}, \& {Mitchell}}]{albrecht2007}
{Albrecht}, S., {Reffert}, S., {Snellen}, I., {Quirrenbach}, A., \& {Mitchell},
  D.~S. 2007, \href{http://dx.doi.org/10.1051/0004-6361:20077953}{\aap, 474,
  565}

\bibitem[{{Albrecht} {et~al.}(2013){Albrecht}, {Setiawan}, {Torres},
  {Fabrycky}, \& {Winn}}]{albrecht2013}
{Albrecht}, S., {Setiawan}, J., {Torres}, G., {Fabrycky}, D.~C., \& {Winn},
  J.~N. 2013, \href{http://dx.doi.org/10.1088/0004-637X/767/1/32}{\apj, 767,
  32}

\bibitem[{{Albrecht} {et~al.}(2011){Albrecht}, {Winn}, {Carter}, {Snellen}, \&
  {de Mooij}}]{albrecht2011}
{Albrecht}, S., {Winn}, J.~N., {Carter}, J.~A., {Snellen}, I.~A.~G., \& {de
  Mooij}, E.~J.~W. 2011,
  \href{http://dx.doi.org/10.1088/0004-637X/726/2/68}{\apj, 726, 68}

\bibitem[{{Andersen}(1975)}]{andersen1975}
{Andersen}, J. 1975, \aap, 44, 355

\bibitem[{{Bakos} {et~al.}(2010){Bakos}, {Torres}, {P{\'a}l}, {Hartman},
  {Kov{\'a}cs}, {Noyes}, {Latham}, {Sasselov}, {Sip{\H o}cz}, {Esquerdo},
  {Fischer}, {Johnson}, {Marcy}, {Butler}, {Isaacson}, {Howard}, {Vogt},
  {Kov{\'a}cs}, {Fernandez}, {Mo{\'o}r}, {Stefanik}, {L{\'a}z{\'a}r}, {Papp},
  \& {S{\'a}ri}}]{bakos2010}
{Bakos}, G.~{\'A}., {Torres}, G., {P{\'a}l}, A., {et~al.} 2010,
  \href{http://dx.doi.org/10.1088/0004-637X/710/2/1724}{\apj, 710, 1724}

\bibitem[{{Barnes} {et~al.}(2011){Barnes}, {Linscott}, \&
  {Shporer}}]{barnes2011}
{Barnes}, J.~W., {Linscott}, E., \& {Shporer}, A. 2011,
  \href{http://dx.doi.org/10.1088/0067-0049/197/1/10}{\apjs, 197, 10}

\bibitem[{{Claret}(2000)}]{claret2000}
{Claret}, A. 2000, \aap, 363, 1081

\bibitem[{{Claret}(2004)}]{claret2004b}
---. 2004, \href{http://dx.doi.org/10.1051/0004-6361:20040470}{\aap, 424, 919}

\bibitem[{{Claret} {et~al.}(2010){Claret}, {Torres}, \& {Wolf}}]{claret2010}
{Claret}, A., {Torres}, G., \& {Wolf}, M. 2010,
  \href{http://dx.doi.org/10.1051/0004-6361/200913942}{\aap, 515, A4}

\bibitem[{{Clausen} \& {Gronbech}(1977)}]{clausen1977}
{Clausen}, J.~V., \& {Gronbech}, B. 1977, \aap, 58, 131

\bibitem[{{De Cat} \& {Aerts}(2002)}]{decat2002}
{De Cat}, P., \& {Aerts}, C. 2002,
  \href{http://dx.doi.org/10.1051/0004-6361:20021068}{\aap, 393, 965}

\bibitem[{{Eggleton} \& {Kiseleva-Eggleton}(2001)}]{eggleton2001}
{Eggleton}, P.~P., \& {Kiseleva-Eggleton}, L. 2001,
  \href{http://dx.doi.org/10.1086/323843}{\apj, 562, 1012}

\bibitem[{{Fabrycky} {et~al.}(2007){Fabrycky}, {Johnson}, \&
  {Goodman}}]{fabrycky2007b}
{Fabrycky}, D.~C., {Johnson}, E.~T., \& {Goodman}, J. 2007,
  \href{http://dx.doi.org/10.1086/519075}{\apj, 665, 754}

\bibitem[{{Fabrycky} \& {Tremaine}(2007)}]{fabrycky2007}
{Fabrycky}, D., \& {Tremaine}, S. 2007,
  \href{http://dx.doi.org/10.1086/521702}{\apj, 669, 1298}

\bibitem[{{Feast}(1954)}]{feast1954}
{Feast}, M.~W. 1954, \mnras, 114, 246

\bibitem[{{Gaposchkin}(1955)}]{gaposchkin1955}
{Gaposchkin}, S. 1955, \mnras, 115, 391

\bibitem[{{Gillon} {et~al.}(2011){Gillon}, {Jehin}, {Magain}, {Chantry},
  {Hutsem{\'e}kers}, {Manfroid}, {Queloz}, \& {Udry}}]{gillion2011}
{Gillon}, M., {Jehin}, E., {Magain}, P., {et~al.} 2011,
  \href{http://dx.doi.org/10.1051/epjconf/20101106002}{in European Physical
  Journal Web of Conferences, Vol.~11, European Physical Journal Web of
  Conferences}, 6002

\bibitem[{{Gray}(2005)}]{gray2005}
{Gray}, D.~F. 2005, {The Observation and Analysis of Stellar Photospheres,
  3$^{\rm rd}$ Ed.} (ISBN 0521851866, Cambridge University Press)

\bibitem[{{Hosokawa}(1953)}]{hosokawa1953}
{Hosokawa}, Y. 1953, \pasj, 5, 88

\bibitem[{{Hut}(1981)}]{hut1981}
{Hut}, P. 1981, \aap, 99, 126

\bibitem[{{Kaufer} {et~al.}(1999){Kaufer}, {Stahl}, {Tubbesing},
  {N{\o}rregaard}, {Avila}, {Francois}, {Pasquini}, \& {Pizzella}}]{kaufer1999}
{Kaufer}, A., {Stahl}, O., {Tubbesing}, S., {et~al.} 1999, The Messenger, 95, 8

\bibitem[{{Kozyreva} \& {Bagaev}(2009)}]{kozyreva2009}
{Kozyreva}, V.~S., \& {Bagaev}, L.~A. 2009,
  \href{http://dx.doi.org/10.1134/S1063773709070056}{Astronomy Letters, 35,
  483}

\bibitem[{{Kupka} {et~al.}(1999){Kupka}, {Piskunov}, {Ryabchikova}, {Stempels},
  \& {Weiss}}]{kupka1999}
{Kupka}, F., {Piskunov}, N., {Ryabchikova}, T.~A., {Stempels}, H.~C., \&
  {Weiss}, W.~W. 1999, \href{http://dx.doi.org/10.1051/aas:1999267}{\aaps, 138,
  119}

\bibitem[{{Lai}(2012)}]{lai2012}
{Lai}, D. 2012,
  \href{http://dx.doi.org/10.1111/j.1365-2966.2012.20893.x}{\mnras, 423, 486}

\bibitem[{{Lehmann} {et~al.}(2013){Lehmann}, {Southworth}, {Tkachenko}, \&
  {Pavlovski}}]{lehman2013}
{Lehmann}, H., {Southworth}, J., {Tkachenko}, A., \& {Pavlovski}, K. 2013,
  \href{http://dx.doi.org/10.1051/0004-6361/201321400}{\aap, 557, A79}

\bibitem[{{Levrard} {et~al.}(2007){Levrard}, {Correia}, {Chabrier}, {Baraffe},
  {Selsis}, \& {Laskar}}]{levrard2007}
{Levrard}, B., {Correia}, A.~C.~M., {Chabrier}, G., {et~al.} 2007,
  \href{http://dx.doi.org/10.1051/0004-6361:20066487}{\aap, 462, L5}

\bibitem[{{Mazeh} \& {Shaham}(1979)}]{mazeh1979}
{Mazeh}, T., \& {Shaham}, J. 1979, \aap, 77, 145

\bibitem[{{Naoz} {et~al.}(2013){Naoz}, {Farr}, {Lithwick}, {Rasio}, \&
  {Teyssandier}}]{naoz2013}
{Naoz}, S., {Farr}, W.~M., {Lithwick}, Y., {Rasio}, F.~A., \& {Teyssandier}, J.
  2013, \href{http://dx.doi.org/10.1093/mnras/stt302}{\mnras, 431, 2155}

\bibitem[{{Pavlovski} {et~al.}(2011){Pavlovski}, {Southworth}, \&
  {Kolbas}}]{pavlovski2011}
{Pavlovski}, K., {Southworth}, J., \& {Kolbas}, V. 2011,
  \href{http://dx.doi.org/10.1088/2041-8205/734/2/L29}{\apjl, 734, L29}

\bibitem[{{Petrie}(1953)}]{petrie1953}
{Petrie}, R.~M. 1953, Publications of the Dominion Astrophysical Observatory
  Victoria, 9, 297

\bibitem[{{Philippov} \& {Rafikov}(2013)}]{philippov2013}
{Philippov}, A.~A., \& {Rafikov}, R.~R. 2013,
  \href{http://dx.doi.org/10.1088/0004-637X/768/2/112}{\apj, 768, 112}

\bibitem[{{Reisenberger} \& {Guinan}(1989)}]{reisenberger1989}
{Reisenberger}, M.~P., \& {Guinan}, E.~F. 1989,
  \href{http://dx.doi.org/10.1086/114972}{\aj, 97, 216}

\bibitem[{{Rogers} {et~al.}(2012){Rogers}, {Lin}, \& {Lau}}]{rogers2012}
{Rogers}, T.~M., {Lin}, D.~N.~C., \& {Lau}, H.~H.~B. 2012, ArXiv,
  \href{http://arxiv.org/abs/1209.2435}{{\sffamily arXiv:1209.2435
  [astro-ph.SR]}}

\bibitem[{{Rogers} {et~al.}(2013){Rogers}, {Lin}, {McElwaine}, \&
  {Lau}}]{rogers2013}
{Rogers}, T.~M., {Lin}, D.~N.~C., {McElwaine}, J.~N., \& {Lau}, H.~H.~B. 2013,
  \href{http://dx.doi.org/10.1088/0004-637X/772/1/21}{\apj, 772, 21}

\bibitem[{{Shakura}(1985)}]{shakura1985}
{Shakura}, N.~I. 1985, Soviet Astronomy Letters, 11, 224

\bibitem[{{Szab{\'o}} {et~al.}(2011){Szab{\'o}}, {Szab{\'o}}, {Benk{\H o}},
  {Lehmann}, {Mez{\H o}}, {Simon}, {K{\H o}v{\'a}ri}, {Hodos{\'a}n},
  {Reg{\'a}ly}, \& {Kiss}}]{szabo2011}
{Szab{\'o}}, G.~M., {Szab{\'o}}, R., {Benk{\H o}}, J.~M., {et~al.} 2011,
  \href{http://dx.doi.org/10.1088/2041-8205/736/1/L4}{\apjl, 736, L4}

\bibitem[{{Torres} {et~al.}(2010){Torres}, {Andersen}, \&
  {Gim{\'e}nez}}]{torres2010}
{Torres}, G., {Andersen}, J., \& {Gim{\'e}nez}, A. 2010,
  \href{http://dx.doi.org/10.1007/s00159-009-0025-1}{\aapr, 18, 67}

\bibitem[{{Triaud} {et~al.}(2013){Triaud}, {Hebb}, {Anderson}, {Cargile},
  {Collier Cameron}, {Doyle}, {Faedi}, {Gillon}, {Gomez Maqueo Chew},
  {Hellier}, {Jehin}, {Maxted}, {Naef}, {Pepe}, {Pollacco}, {Queloz},
  {S{\'e}gransan}, {Smalley}, {Stassun}, {Udry}, \& {West}}]{triaud2013}
{Triaud}, A.~H.~M.~J., {Hebb}, L., {Anderson}, D.~R., {et~al.} 2013,
  \href{http://dx.doi.org/10.1051/0004-6361/201219643}{\aap, 549, A18}

\bibitem[{{van Houten}(1950)}]{houten1950}
{van Houten}, C.~J. 1950, Annalen van de Sterrewacht te Leiden, 20, 223

\bibitem[{{van Leeuwen}(2007)}]{vanleeuwen2007}
{van Leeuwen}, F. 2007,
  \href{http://dx.doi.org/10.1051/0004-6361:20078357}{\aap, 474, 653}

\bibitem[{{Waelkens}(1991)}]{waelkens1991}
{Waelkens}, C. 1991, \aap, 246, 453

\bibitem[{{Yakut} {et~al.}(2007){Yakut}, {Aerts}, \& {Morel}}]{yakut2007}
{Yakut}, K., {Aerts}, C., \& {Morel}, T. 2007,
  \href{http://dx.doi.org/10.1051/0004-6361:20065506}{\aap, 467, 647}

\bibitem[{{Yakut} {et~al.}(2014){Yakut}, {Aerts}, \& {Morel}}]{yakut2014}
---. 2014, \href{http://dx.doi.org/10.1051/0004-6361/20065506e}{\aap, 562, C2}

\bibitem[{{Zhou} \& {Huang}(2013)}]{zhou2013}
{Zhou}, G., \& {Huang}, C.~X. 2013,
  \href{http://dx.doi.org/10.1088/2041-8205/776/2/L35}{\apjl, 776, L35}

\end{thebibliography}
\end{document}